\documentclass[preprint,aps,nofootinbib]{revtex4}
\pdfoutput=1
\usepackage{graphicx}
\usepackage{epstopdf}    
\usepackage{dcolumn}
\usepackage{amsfonts}
\usepackage{amsmath}
\usepackage{amssymb}
\usepackage{bm}
\usepackage{color}
\usepackage{comment}
\usepackage{epsf}
\usepackage{float}
\usepackage{enumitem}

\newcolumntype{L}{>{\centering\arraybackslash}m{3cm}}
	
	\newcommand{\bqa}{\begin{eqnarray}}
	\newcommand{\eqa}{\end{eqnarray}}
	\newcommand{\bwt}{\begin{widetext}}
	\newcommand{\ewt}{\end{widetext}}

	\newcommand{\nn}{\nonumber \\}

\newcommand{\edc}{\end{document}}
\newcommand{\bb} {\begin{thebibliography}}
\newcommand{\eb}{\end{thebibliography}}
\newcommand{\bc}{\begin{center}}
\newcommand{\ec}{\end{center}}

\newcommand{\be}{\begin{equation}}
\newcommand{\ee}{\end{equation}\normalsize}
\newcommand{\bea}{\begin{eqnarray}}
\newcommand{\eea}{\end{eqnarray}}
\newcommand{\ba}{\begin{array}{l}   }
\newcommand{\lab}[1]{\label{#1}}
\newcommand{\ea}{\end{array}}

\newcommand{\dsfrac}{\displaystyle\frac}
\newcommand{\ds} {\displaystyle}

\newcommand{\re}[1]{(\ref{#1})}

\newcommand{\ci}{\cite}

\newcommand{{\vergul}}{  ,}

\newcommand{\veps}{\varepsilon }

\newcommand{\gamtil}{\widetilde{\gamma}}

\begin{document}
\draft
\title{Spin-gapped magnets with weak anisotropies
II: Effects of exchange and Dzyaloshinsky-Moriya  anisotropies on thermodynamic characteristics}

	\author{Abdulla Rakhimov$^1$, Asliddin Khudoyberdiev$^2$,  B. Tanatar$^1$ }
	\affiliation{$^1$Department of Physics, Bilkent University, Bilkent, 06800 Ankara, Turkey\\
	$^2$Institute of Nuclear Physics, Tashkent 100214, Uzbekistan}

		\date{\today}
		\begin{abstract}
We study the modification of low temperature properties of quantum magnets such as magnetization,
heat capacity, energy spectrum, and densities of condensed and noncondensed quasiparticles (triplons)
due to anisotropies in the framework of mean-field based approach. We show that in contrast to
exchange anisotropy (EA) interaction, Dzyaloshinsky-Moriya (DM) interaction modifies the physics dramatically.
Particularly, it changes the sign of the anomalous density in the whole range of
temperatures. Its critical behavior is slightly modified also by the EA. We have found that the shift of the critical
temperature of phase transition (or crossover caused by DM interaction) is positive and significant.
Using the experimental data
on the magnetization of the compound TlCuCl$_3$, we have found optimal values
for the strengths of EA  and DM interactions. The spectrum of the energy of low lying
excitations has also been investigated and found to develop a linear dispersion similar to Goldstone mode
with a negligibly small anisotropy gap.
\end{abstract}

		\maketitle

\section{Introduction}
\label{sec:1}
Presently, it is well established that there is a class of quantum magnets whose low temperature properties could be
described within the paradigm of Bose-Einstein condensation (BEC) of quasiparticles referred as triplons\cite{Zapf}.
Experimentally this is confirmed by studying the critical exponents as well as the magnetic excitation spectrum of
such compounds at low temperatures. A good example is the critical exponent $\phi$, associated with phase
boundary $H_c(T)$, that divides the paramagnetic and field induced canted XY antiferromagnetic phase of
several quantum magnets, $H_c(T)-H_c(0)\propto T^\phi$. This exponent approaches its expected value of 3/2,
which is typical for a system with BEC, when the window of low temperatures is rather reduced\cite{shermansound}.
Recent experimental investigations were conducted by Zhou {\it et al.}\cite{tanakanew} for the entire magnetization
process of TlCuCl$_3$ up to the magnetic field of 100\,T at  temperature 2\,K. They also analyzed magnetic
field-temperature phase boundary dependence around the critical fields $H_{c1}$ and $H_{c2}$ and concluded that
for both critical phase boundaries the critical exponents are $\phi\approx 3/2$.  Another experimental evidence is
offered by the properties of the excitation spectrum in the BEC state which has been theoretically predicted to be a
gapless Goldstone mode associated with the spontaneous breaking of rotational symmetry by the staggered order.
Thus, the presence of a spin-wave like mode with a linear mode dispersion, $E_k\sim ck$, is a convincing signal
for the existence of BEC in this class of quantum magnets\cite{{matsumotoprl}}. Therefore, one may conclude that at low temperatures thermodynamic properties of such materials are determined mainly (but not only) by the condensation of triplons\cite{yukalovtriplon}.

Theoretically, the system of triplons can be described by the following effective Hamiltonian:

\bea
H_{iso}&=\int d^3r\left[\psi^+(\vec{r}) (\hat{K}-\mu) \psi(\vec{r})+ \frac{U}{2} (\psi^+(\vec{r})\psi(\vec{r}))^2\right]\, ,\label{1b}
\eea

where $\psi(r)$ is the bosonic field operator, $\hat{K}$ is the kinetic energy operator which defines the bare triplon dispersion $\veps_k$ in momentum space and $U$ is the strength of contact interaction describing a strong
short-range triplon-triplon repulsion. The Hamiltonian in Eq.\,(\ref{1b}) is formally the same as used for BEC of
atomic gases\cite{andersen}. However, there is a small difference in the
strategy. In tasks related to atomic Bose gases the number of particles $N$ is assumed to be fixed, while
the chemical potential $\mu(N,T)$ is to be calculated say, by the relation
$N\sim\sum_k1/[e^{\beta(\veps_k-\mu)}-1]$, where $\beta$ is the inverse temperature.
As to the triplon gas, the chemical potential in Eq.\,(\ref{1b}) characterizes an additional direct contribution to the
triplon energy due to the external magnetic field $H$, giving $\mu=g\mu_B(H-H_c)$ where $g$ is the electron
Land{\'e} factor,
$\mu_B=0.672$ KT$^{-1}$ is the Bohr magneton and $H_c$ is the critical magnetic field which defines the gap
 $\Delta_{ST}=g\mu_BH_c$ between singlet and triplet states.
In the field induced BEC, $\mu$ is assumed to be an input parameter, from which the total number of triplons
can be calculated.
Moreover, for homogenous atomic gases one may use simple quadratic bare dispersion $\veps_k=k^2/2m$ with a good accuracy, while for spin-gapped quantum magnets a more complicated form of bare dispersion
is needed.\cite{shermansound,cavadaniepj,matsumotoPRB,misguich,wangprb2014}

It is well known that the Hamiltonian in Eq.\,(\ref{1b}) leads to a gapless Bogoliubov dispersion,
$E_k=\sqrt{\veps_k}\sqrt{\veps_k+2U\rho}\approx ck+\mathcal{O}(k^3)$ at low temperatures, with density $\rho=N/V$
and sound velocity $c$. However, low frequency electron spin resonance (ESR)
measurements on some materials, such as TlCuCl$_3$\cite{kolezuk,glazkov},
(C$_4$H$_{12}$N$_2$)(Cu$_2$Cl$_6$)\ci{glazkov2012}, Cs$_2$CuCl$_4$\ci{povarov2011}, DTN\ci{zvyagin2007}
gave evidence for a tiny spin gap. The origin of this gap is due to exchange anisotropy (EA) or
Dzyaloshinsky-Moriya (DM) interactions, which should be taken into account in the theoretical description,
and  particularly, in the effective model Hamiltonian\cite{Miyahara}. A simpler extended Hamiltonian such
as Eq.\,(\ref{1b}) including EA and DM interactions was proposed by Sirker {\it et al.}\cite{Sirker1}
\be
\ba
H_{aniz}=\int d^3r\{\psi^+(\vec{r}) (\hat{K}-\mu) \psi(\vec{r})+ \dsfrac{U}{2} (\psi^+(\vec{r})\psi(\vec{r}))^2+\dsfrac{\gamma}{2}[\psi^+(\vec{r})\psi^+(\vec{r}) + \psi(\vec{r})\psi(\vec{r})]\\ +i\gamma'[\psi(\vec{r})-\psi^+(\vec{r})]\}
\label{Hbde}
\ea
\ee
where $\gamma$ and $\gamma'$ are interaction strengths of EA and DM interactions, respectively
($\gamma\geq 0$, $\gamma'\geq 0$). Thus, once the Hamiltonian is given, one first separates fluctuations as
$\psi=\xi\sqrt{\rho_0}+\tilde{\psi}$, where $\xi=e^{i\Theta}$ and $\rho_0$ are the phase of the condensate wave
function and its magnitude, respectively; and then introducing second quantization, $\tilde{\psi}(\vec{r})=\sum_ke^{i\vec{k}\vec{r}}a_k$,  $\tilde{\psi}^+(\vec{r})=\sum_ke^{-i\vec{k}\vec{r}}a_k^+$, makes an attempt to diagonalize the Hamiltonian $H$ with respect to creation ($a^+$) and annihilation ($a$) operators. As a result, analytical expressions for quasiparticle (bogolon) dispersion $E_k$ and some other quantities may be obtained. In the present work we shall take into account anomalous averages $\sigma=\sum_{k}\sigma_{k}=\frac{1}{2}\sum_k
     \left(\langle a_{k}a_{-k}\rangle+\langle a_{k}^{\dag}a_{-k}^{\dag}\rangle\right) $ ($\sigma$-anomalous density) based on Hartree-Fock-Bogoliubov approach, which was neglected in Ref.\,  \cite{Sirker1}.
     This allows one to obtain continuous magnetization across the BEC transition, which would be
     discontinuous otherwise, in the so-called Hartree-Fock-Popov (HFP) approximation with $\sigma=0$\ci{ourANN}.

In order to get more information about thermodynamics of the system we exploit the grand canonical thermodynamic
potential $\Omega$, which may be evaluated in the path integral formalism\cite{andersen,cooper,klbookfi,ouryee}.
 This will be convenient to study the modification of the
condensate wave function, entropy $S=-(\partial\Omega/\partial T)$, heat capacity $C_H=T(\partial S/\partial T)$, magnetization $M=-(\partial\Omega/\partial H)$, and possibly other physical quantities due to anisotropies.

In our previous work\cite{ourpart1} we have derived an explicit expression for $\Omega$ of a homogenous
system of bosons, described by the Hamiltonian in Eq.\,(\ref{Hbde}). Minimization of thermodynamic potential with respect to the phase $\xi$ and condensate fraction $\rho_0$, together with the requirement of dynamical stability of BEC led to following conclusions (see Table 1 of  Ref.\, \cite{ourpart1}).

(a) The condensate has a definite phase\cite{barnett}, which is independent of temperature or magnetic field.

(b) The phase angle $\Theta$ may have only discrete values, namely $\Theta=\pi n$ and $\Theta=\pi/2+2\pi n$ ($n=0, \pm 1 , \pm 2...$) for an equilibrium system of bosons without and with DM interaction, respectively.

(c) The presence of a weak DM interaction even with a tiny strength smears out the phase transition from BEC to
normal phase into a crossover, i.e, the condensate fraction may vanish only asymptotically by increasing the
temperature. Besides, the DM interaction fixes the direction
of staggered magnetization, predicted by Matsumoto {\it et al.}\ci{matsumoto2008}, based on symmetry
considerations.

In the present work we shall study the modification of some physical observables due to EA and DM anisotropies given by Eq.\,(\ref{Hbde}).

The rest of this paper is organized as follows. In Section II we discuss the properties of main equations
of the present approach. In Section III we analyze the role of anisotropies for the thermodynamic parameters such as anomalous density, self-energies, magnetization and heat capacity. We compare our theoretical results with experimental ones for the TlCuCl$_3$ compound in Section IV and summarize our main
results in Section V.

Throughout the paper we adopt the units $k_{B}\equiv1$ for the Boltzmann constant, $\hbar\equiv1$
for the Planck constant, and $V\equiv1$ for the unit cell volume.
In these units the energies are
measured in Kelvin (K), the mass $m$ is expressed in K$^{-1}$, the magnetic susceptibility
$\chi$ for the magnetic fields measured in Tesla (T) has the units of \rm{K/T$^2$}, while the momentum
and specific heat $C_H$ are dimensionless. Particularly, the Bohr magneton is
$\mu_{B}={\hbar e}/{2m_{0}c}=0.671668$ \rm{K/T}, where $m_{0}$ is the free electron mass, and $e$ is the fundamental charge.

\section{Properties of main equations for self energies}

One of the main quantities to describe the low temperature properties of ultracold bosonic systems is the dispersion relation for quasiparticles, which is supposed to be written as $E_k=\sqrt{\varepsilon_k+X_1}\sqrt{\varepsilon_k+X_2}$, in general. Here $\varepsilon_k$ is the bare dispersion of triplons\cite{23 our aniz} and the quantities $X_{1,2}$ are related to the ordinary normal $\Sigma_{n}$,
and anomalous $\Sigma_{an}$, self-energies as follows $X_{1,2}=\Sigma_{n}\pm \Sigma_{an}-\mu $.
The self-energies $X_{1,2}$ and the condensate fraction are the solutions to the following equations\ci{ourpart1}:
 \begin{subequations}
	\begin{align}
X_1&= 2U\rho + U\sigma -\mu + \frac{U\rho_0(\xi^2+ \bar{\xi}^2)}{2}+\gamma +\frac{2\gamma'^2 D_1}{X_2^2} \label{eq:X10}\\
X_2&=2U\rho - U\sigma -\mu - \frac{U\rho_0(\xi^2+ \bar{\xi}^2)}{2}-\gamma -\frac{2\gamma'^2 D_2}{X_2^2}
\label{eq:X20}\\
\frac{\partial\Omega}{\partial\rho_0}
&=
\cos2\Theta (U\sigma + \gamma) + U (\rho_0 +2\rho_1)-\mu -\frac{\gamma' \sin\Theta}{\sqrt{\rho_0}} =0
\label{eq:rho0}
\end{align}
\end{subequations}
where
\begin{subequations}
	\begin{align}
	A_1'&= \frac{\partial A}{\partial X_1}= \frac{1}{8}\sum_{k}\frac{(E_k W_k' + 4W_k)}{E_k}\\
    A_2'&= \frac{\partial A}{\partial X_2}=\frac{1}{8}\sum_{k}\frac{(\veps_k + X_1)^2(E_k W_k' - 4W_k)}{E_k^3}\\
    B_1'&= \frac{\partial B}{\partial X_1} = \frac{1}{8}\sum_{k}\frac{(\veps_k + X_2)^2(E_k W_k' - 4W_k)}{E_k^3}\\
   D_1&=\frac{A_1'}{\bar{D}}; \quad D_2=\frac{B_1'}{\bar{D}}; \quad \bar{D}=A_1'^2 - A_2'B_1'\\
   W_k& =\frac{\coth(\beta E_k/2)}{2} ;     \quad W_k' =\beta (1-4W_k^2)=\frac{-\beta}{\sinh^2(\beta E_k/2)}\, .
    \label{a123}
	\end{align}
\end{subequations}
In the above equations $A=\rho_1-\sigma$,  $B=\rho_1+\sigma$ and the normal $\rho_1$ and anomalous
$\sigma$ densities are given below. In the Hartree-Fock-Bogoliubov approximation these self-energies play an
essential role. Thus, we first study their properties and then evaluate physical observables under consideration.
For simplicity, we rewrite Eqs.\,(2.3) and (2.4) separately for the cases with ( $\gamma'=0$) and
without ($\gamma' \neq0$) DM interactions, taking into account that for these cases $\xi=\pm 1$
and $\xi=i$, respectively.

\subsection{mode 1: $\gamma'=0,  \gamma\neq0, \xi =1$}
This mode corresponds to the case when only EA is present. Here we have both phases, BEC and  normal,
which are sharply separated by the critical temperature $T_c$ defined by the equation $\rho_0(T=T_c)=0$.
The condensate fraction is given in BEC phase by $\rho_0=(\Delta -2\gamma-U\sigma)/U$, where $\Delta$
is the solution of the algebraic equation
 \bea
 \Delta=\mu + 2U(\sigma-\rho_1) + 5\gamma=U(\rho_0+\sigma)+2\gamma
 \label{eq:Delta}
 \eea
where the normal $\rho_1$ and anomalous $\sigma$ densities are given by following general expressions
\begin{subequations}
	\begin{align}
	\rho_1 &=  \sum_k\left[\frac{W_k(\veps_k+X_1/2 +X_2/2)}{E_k} -\frac{1}{2} \right] \equiv\sum_k \rho_{1k} \label{eq:17a}\\
	\sigma & =\frac{(X_2-X_1)}{2} \sum_k \frac{W_k}{E_k}  \equiv\sum_k \sigma_{k}
	\label{eq:rho1sig}
	\end{align}
\end{subequations}
 with $X_1=2\Delta$, $X_2=2\gamma$, $E_k=\sqrt{(\veps_k+2\Delta)(\veps_k+2\gamma})$, $\rho=\rho_0+\rho_1$.

In the normal phase $\rho_0(T>T_c)=0$,  the self-energies $X_1$, $X_2$ in the  dispersion relation $E_k\equiv\omega_k=\sqrt{(\veps_k+X_1)(\veps_k+X_2)}$ are given as
\bea
X_{1,2}(T>T_c)=2U\rho-\mu \pm(U\sigma +\gamma)
\label{eq:49}
\eea
where the total triplon density is
\bea
\rho(T>T_c) = \sum_{k}\frac{1}{e^{\beta\omega_k}-1}\, .
\label{eq:50}
\eea
Explicit expressions for other quantities are moved to the Appendix  for convenience.

Note that our mean-field based main equations are rather general leading to well known approximations used in the literature for the isotropic case, when all anisotropies are neglected. Particularly, one may derive Hartree-Fock-Popov approximation and simple Bogoliubov approximations as follows.
\begin{itemize}
\item{\bf{HFP approximation}}
This is widely used in literature and obtained simply by neglecting $\sigma$ and $\gamma$ in Eq.\,(\ref{eq:Delta})
resulting in the following equation for the condensate fraction
\bea
\rho_0=\rho-\rho_1=\rho-\sum_k\left[
\frac
{
W_k(\veps_k+U\rho_0)
}
{
\sqrt{\veps_k}\sqrt{\veps_k+2U\rho_0}
}
-\frac{1}{2}
\right]\, .
\label{eq:hfp}
\eea
\item{\bf{Bogoliubov approximation}}.
Further, at zero temperature, making formal replacement $\rho_0\rightarrow\rho$ on the right hand side of
Eq.\,(\ref{eq:hfp}) gives
\bea
\frac
{\rho_0}
{\rho}
=1-\frac
{1}
{2\rho}
\sum_k\left[\frac
{\veps_k+U\rho}
{
\sqrt{\veps_k}\sqrt{\veps_k+2U\rho}
}-
1
\right]\, .
\label{bogol}
\eea
For infinite uniform system with $\veps={\vec k}^2/{2m}$, $|k|=0,...\infty$, one may evaluate the momentum
integration in Eq.\,\re{bogol}  to obtain the following well known formula\cite{ouryee,ourANN}
\be
\frac{\rho_0}{\rho}=1-\frac{8\sqrt{\rho a^{3}_{s}}}{3\sqrt{\pi}}
\lab{bogrho0}
\ee
where $a_s=Um/4\pi$  is the s-wave scattering length. Remarkably, the quantum depletion given by the second term
on the right hand side of Eq.\,\re{bogrho0}, as well as the energy dispersion in Eq.\,\re{bogol} $E_k=\sqrt{\veps_k}\sqrt{\veps_k+2U\rho}$ were proposed by Bogoliubov more then
seventy years  ago\ci{bogol} and has been one of the cornerstones of our understanding of
interacting quantum fluids\ci{rect_exper}.
\end{itemize}

\subsection{mode 2: $\gamma'\neq 0, \gamma\neq0, \xi =i$}
Here both EA and DM interactions are present. The main equations for self-energies $X_1$ and $X_2$ are
obtained from Eq.\,(20) of Ref.\, \cite{ourpart1}  by setting
$\xi=i$,
\begin{subequations}
	\begin{align}
X_1&=2U\sigma+2\gamma +\frac{\gamma'}{\sqrt{\rho_0}}+\frac{2\gamma'^2 D_1}{X_2^2}  \label{eq:GX1}\, ,\\
X_2&=2U\rho_0 +\frac{\gamma'}{\sqrt{\rho_0}}-\frac{2\gamma'^2 D_2}{X_2^2}\,. \label{eq:GX2}
\end{align}
\end{subequations}
The equation for the condensate fraction $\rho_0$ may be presented in the following dimensionless
compact form
\bea
r_0^3 +Pr_0 +Q=0\,
\label{eq:r0}
\eea
where we have introduced
$P =-\bar{\sigma} + 2(\bar{\rho_1}-1-\bar{\gamma})$, $ Q=-2{\bar{\gamma}'}/{\sqrt{\rho_{c0}}}$,
$r_0^2=\rho_0/\rho_{c0}  $, $ \bar{\sigma}={\sigma}/{\rho_{c0}}$,  $\bar{\rho_1}={\rho_1}/{\rho_{c0}}$, $\bar{\gamma}={\gamma}/{\mu}$,  $\bar{\gamma}'={\gamma'}/{\mu}$ in which
$\rho_{c0}$ is the critical density of pure BEC,  $\rho_{c0}={\mu}/{2U}$.

In general, one has to solve these three coupled nonlinear algebraic equations for the unknown quantities $X_1$,
$X_2$ and $r_0$ at a given temperature and magnetic field. Clearly, in such cases it is important to guess the initial
values of $X_1(T)$ and $X_2(T)$, since the solutions are not unique. For this purpose it will be
convenient to start from a higher temperature, say $T \approx15$\,K, where $\sigma(T\gg T_c)\approx 0$, $\gamma'^2/X_2^2 \rightarrow 0$ and hence Eqs.\,(2.12) are simplified to
\bea
 Z_1=\frac{\gamma}{\mu}-\frac{Q}{4 r_0}, \quad  Z_2=\frac{r_0^2}{2}-\frac{Q}{4 r_0},
 \label{eq:Z1Z2}
\eea
where $Z_1=X_1/2\mu$ and $Z_2=X_2/2\mu$.\\

\subsubsection{High temperatures}
For a weak EA interaction, $\gamma/\mu\ll 1$ Eqs.\,\re{eq:Z1Z2} coincide with those obtained by
Sirker {\it et al.}\cite{Sirker2} within the HFP approximation with $\sigma=\gamma=0$, and may be solved
easily by inserting $Z_1$, $Z_2$ into Eq.\,\re{eq:r0}, thus by reducing the system of three coupled equations
into one cubic algebraic equation with respect to $r_0$. It is clear that in this regime
Eqs.\,\re{eq:X10} and \re{eq:X20} are simplified as
\bea
X_1(T\gg T_c) \approx X_2(T \gg T_c)= 2U\rho -\mu
\label{eq:55}
\eea
where $\rho=\rho_0+\rho_1$ is the total density of triplons, and $T_c$ is defined as
$(d\rho/dT)|_{T=T_c}=0$, $(d^2\rho/dT^2) |_{T=T_c}\geq 0$ and hence the normal $\Sigma_{n}$ and anomalous
$\Sigma_{an}$ self-energies have the form
 \bea
\Sigma_{n}=\mu+\frac{X_1+X_2}{2}\approx 2U\rho,  \quad   \quad
\Sigma_{an}=\frac{X_1-X_2}{2}  \approx 0\, .
\label{eq:56}
\eea

\begin{figure}[bt]
\begin{minipage}[bt]{0.49\linewidth}
\center{\includegraphics[width=1.1\linewidth]{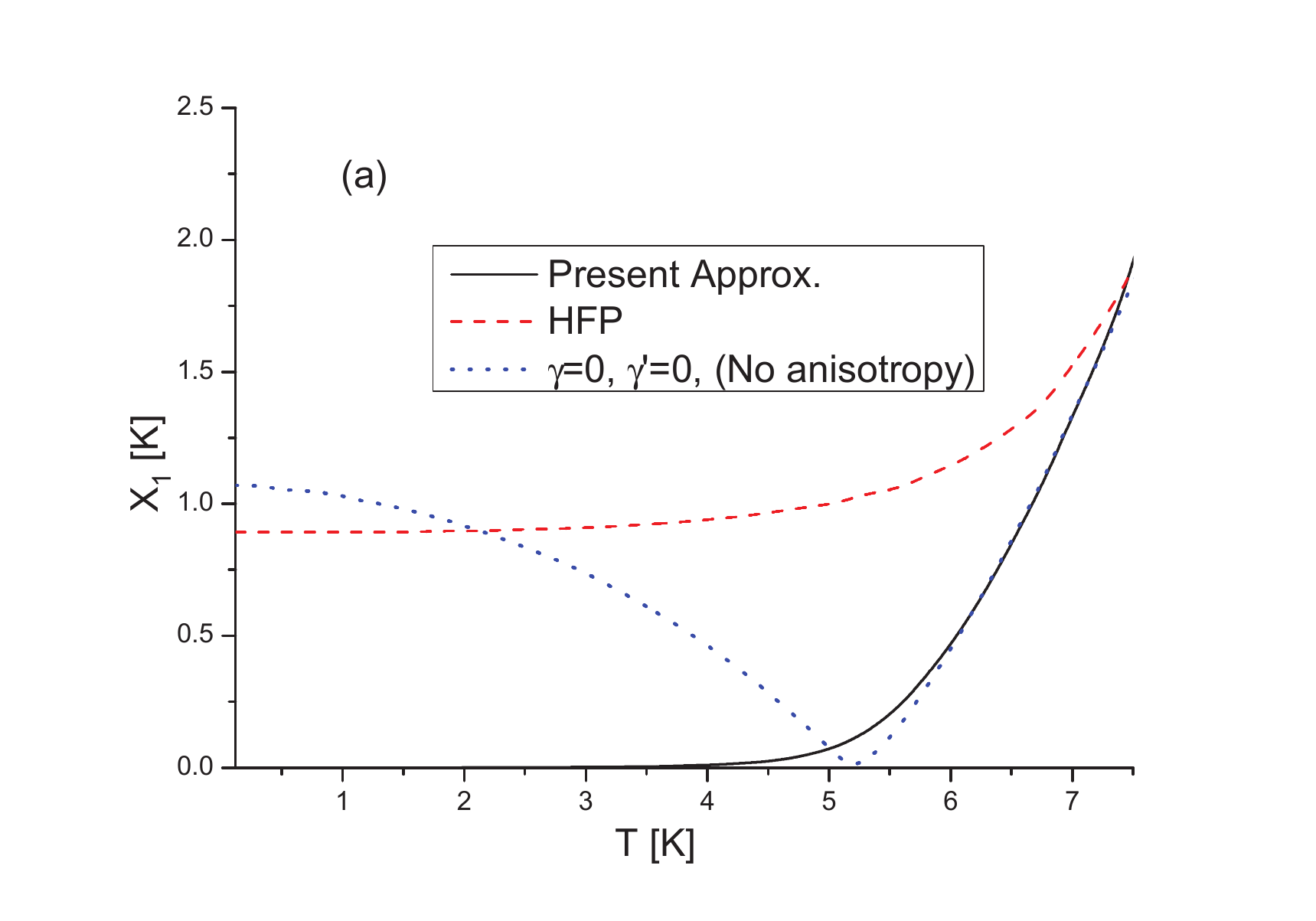} }
\end{minipage}
\hfill
\begin{minipage}[bt]{0.49\linewidth}
\center{\includegraphics[width=1.1\linewidth]{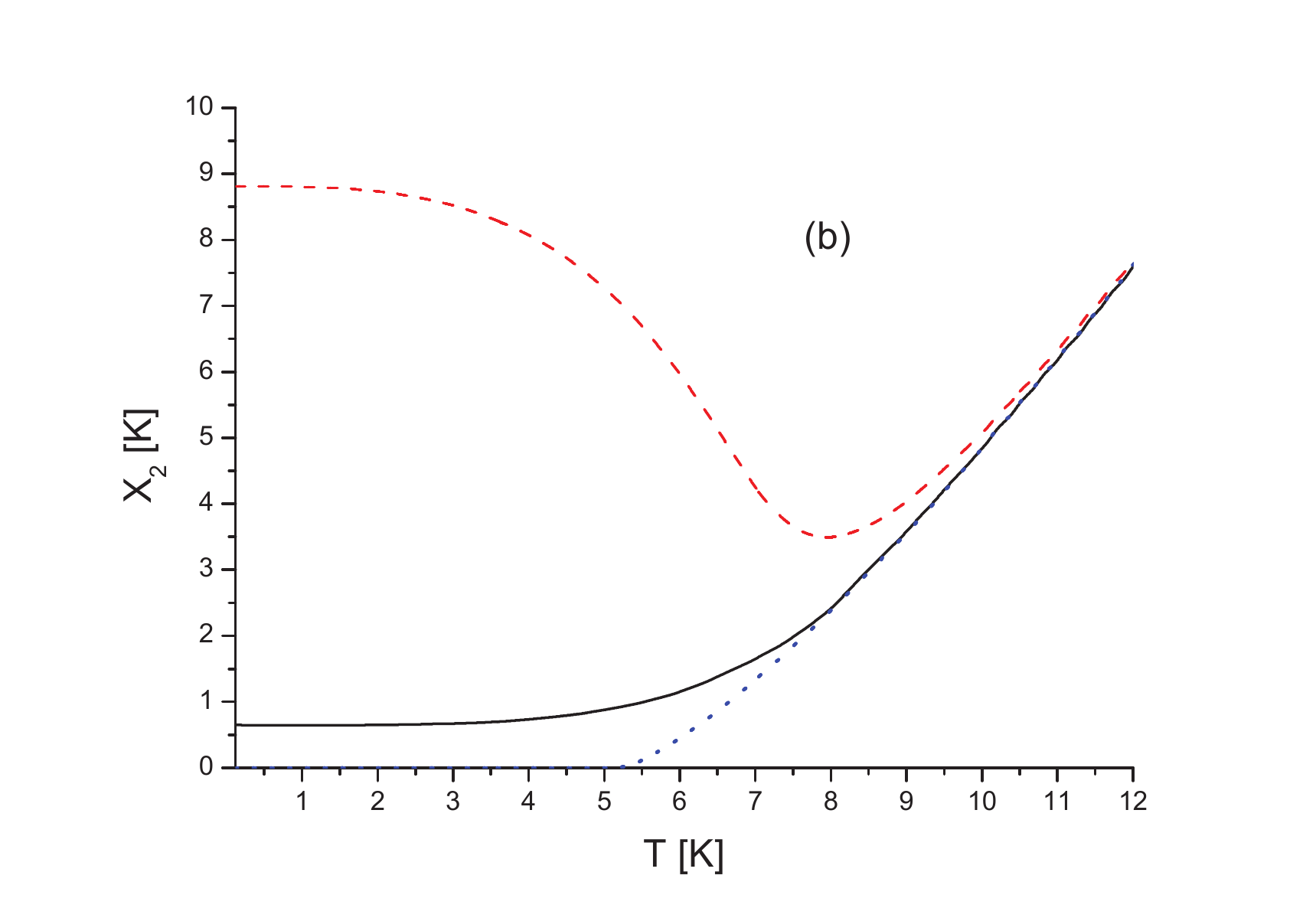} }
\end{minipage}
\medskip
\medskip
\medskip
\begin{minipage}[bt]{0.49\linewidth}
\center{\includegraphics[width=1.1\linewidth]{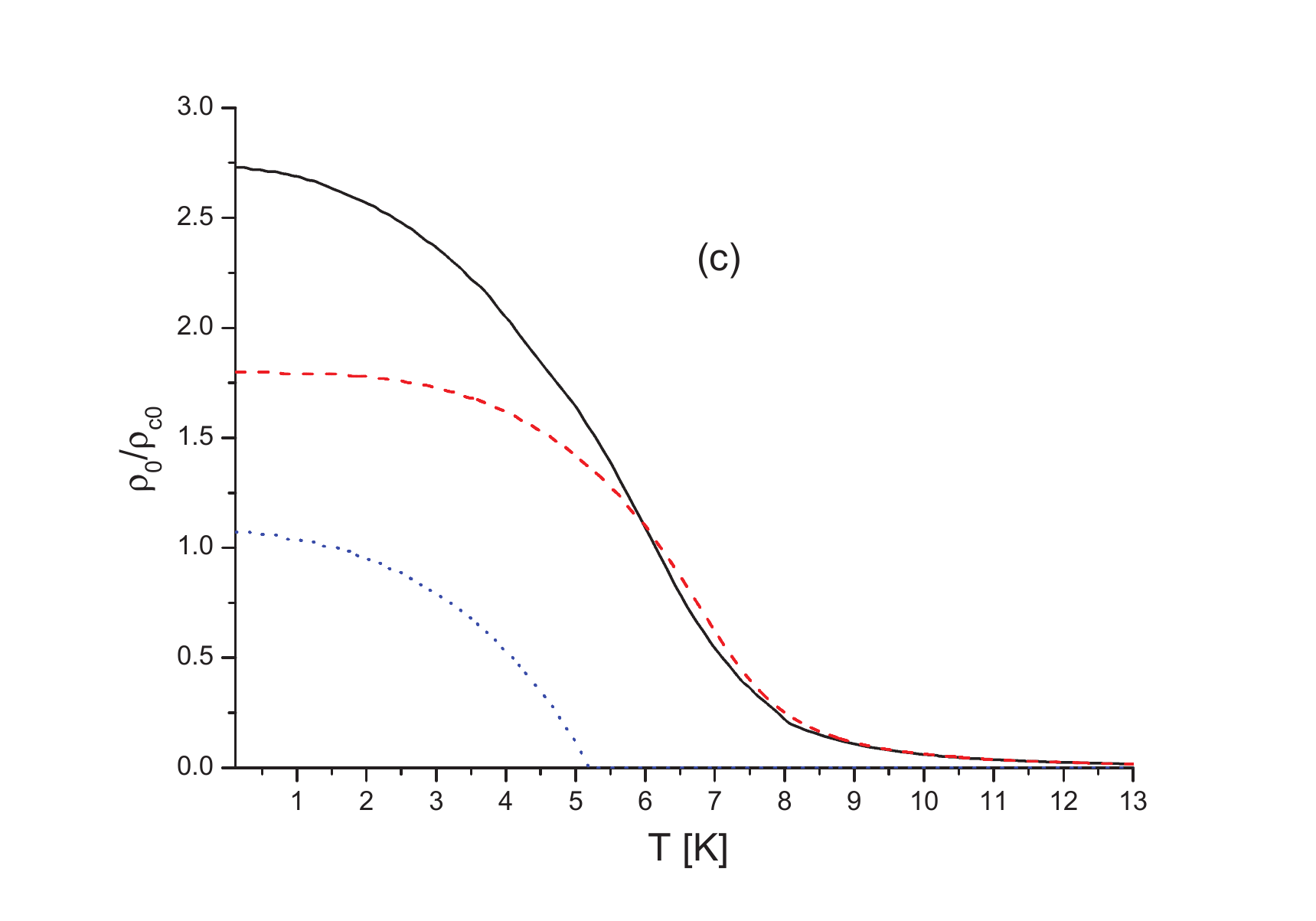}
}
\end{minipage}
\hfill
\begin{minipage}[bt]{0.49\linewidth}
\center{\includegraphics[width=1.1\linewidth]{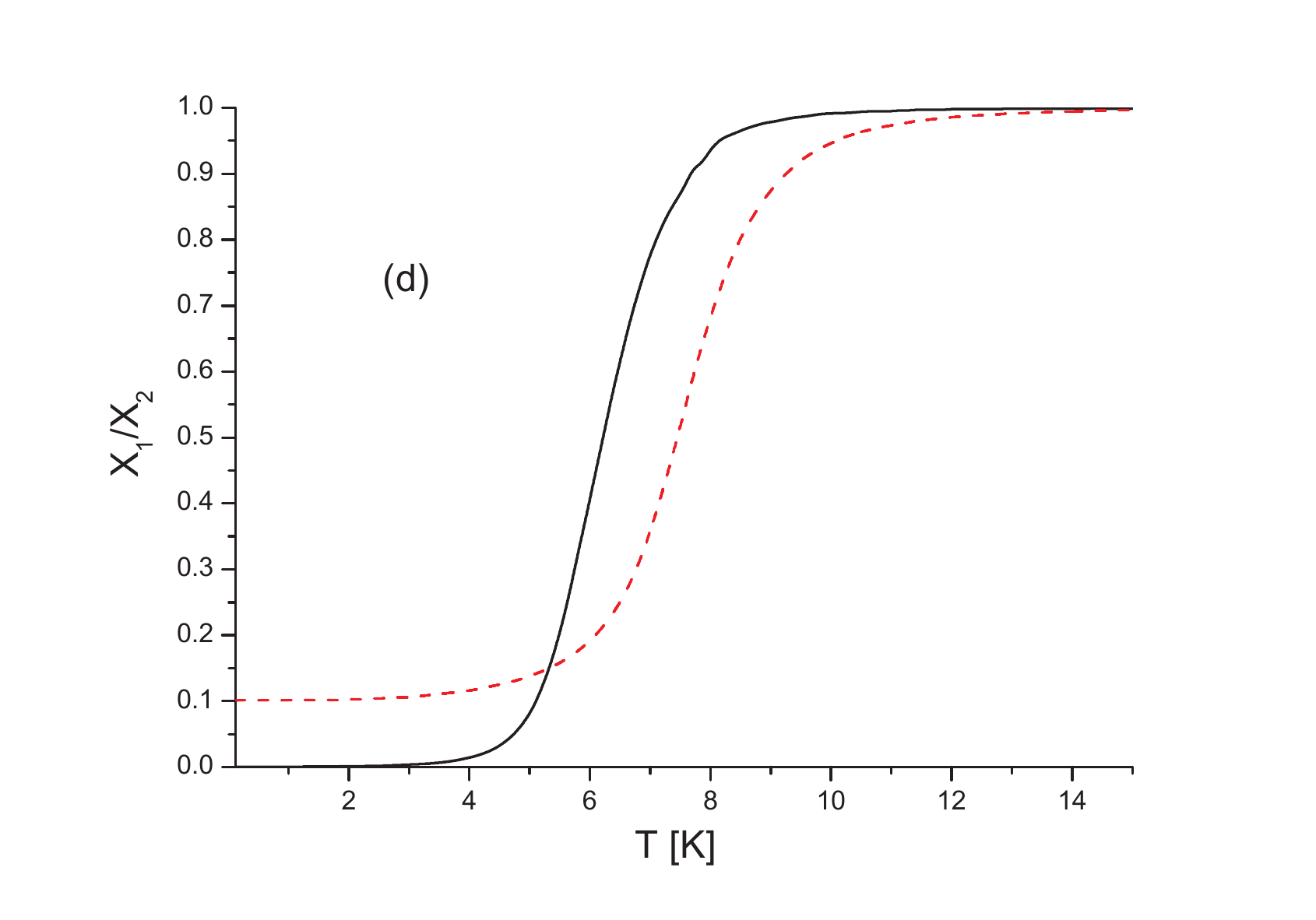}
}
\end{minipage}
\caption
{Physical solutions of Eqs.\,(2.12)  and \re{eq:r0}  with only anisotropic DM interaction for the input parameters
$g=2.06$, $U=315$\,K, $H=8.5$\,T, $\gamma'=0.1$\,K, and $\gamma=0$ as a function of temperature.
The parameters of bare dispersion $\veps_k$ are taken from Ref. [9]; (a), (b) and (c)
represent the self-energies $X_1(T)$, $X_2(T)$ and the condensate fraction
$\rho_0(T)/\rho_{0c}$ $(\rho_{0c}=\mu/2U=0.07)$, respectively, while (d) illustrates the ratio $X_1(T)/X_2(T)$.
Solid and dashed lines correspond to present approximation and that by Ref. [28], corresponding
to the case with formally setting $\sigma=\gamma=(\gamma')^2=0$ in Eqs.\,(2.12) and \re{eq:r0}, respectively. The dotted lines represent isotropic case with $\gamma=\gamma'=0$.}
  \label{FIGX12RHO0X}
\end{figure}

In Fig.\,\ref{FIGX12RHO0X}, we present typical solutions of Eqs.\,(2.12) and \re{eq:r0}
as a function of temperature for $\gamma'=0.1$\,K and $\gamma =0$. It is seen that at high temperatures
$X_1$  and $X_2$ overlap with that of pure BEC with $\gamma=\gamma'=0$ in accordance with
Eq.\,\re{eq:55}. Therefore, the effect of anisotropy on self-energies is negligibly small
at high temperatures. On the other hand, the effect of DM interaction on the condensate
fraction is rather significant, as it is seen from Fig.\,\ref{FIGX12RHO0X}(c).

In fact, since in the presence of DM interaction the parameter $Q$ is finite, Eq.\,\re{eq:r0} does not have a
zero solution, as illustrated in Fig.\,\ref{FIGX12RHO0X}(c). Strictly speaking, at any temperature there exists
a finite condensate fraction. Thus, comparing $\rho_0(T)$ for pure BEC (dotted curve)
with that for the case of DM interaction (solid curve) in Fig.\,\ref{FIGX12RHO0X}(c) one
may conclude that, DM anisotropy smears out BEC transition into a crossover.

\subsubsection{Low temperatures}
Moreover, comparing those curves in Fig.\,\ref{FIGX12RHO0X}(c) at low temperatures one may note that
the DM interaction enhances the condensate fraction significantly.
For example, the condensate fraction at $T=0$ for $\gamma'=0.1$\,K
is nearly 2.7 times larger than that for $\gamma'=0$, corresponding to the isotropic case.

We now discuss the low temperature behavior of self-energies $X_1$ and $X_2$.
As it is seen from Fig.\,\ref{FIGX12RHO0X} in this region in the  approach by  Sirker et al.\ci{Sirker2}
$X_1$ and $X_2$ are nearly of the same order, while in the present approximation $X_1$ is much
smaller than $X_2$, ($X_1/X_2 \approx 10^{-4}$). The main reason of this difference is that in the present
approximation the anomalous density has not been neglected, and besides,
the DM interaction is taken into account up to the second order in the strength.
Now coming back to the main equations for $X_1$ and $X_2$ one may note that,
at low temperatures, $D_1$ in Eqs.\,\re{eq:GX1} given by Eqs.\,(2.4) becomes negligibly small, while $D_2$ in
Eq.\,(\ref{eq:GX2}) remains finite. Thus, Eq.\,(\ref{eq:GX1}) with $\gamma=0$ and the difference $X_2-X_1$
can be written as
\begin{subequations}
	\begin{align}
X_1(T \rightarrow 0) &\approx 2U\sigma +\frac{\gamma'}{\sqrt{\rho_0}} \label{eq:58a}\\
(X_2-X_1)|_{T \rightarrow 0} &\approx 2U(\rho_0-\sigma) -\frac{2\gamma'^2 D_2}{X_2^2}\, .
\label{eq:58b}
\end{align}
\end{subequations}
From Eq.\,(\ref{eq:58a}) it can be immediately seen that, since $\sigma>0$,\footnote{See the next section.}
$X_1(T \rightarrow 0) \neq 0$ when $\gamma' \neq 0$, that is the gap in the quasiparticle dispersion
$ E_k= \sqrt{(\veps_k+X_1)(\veps_k+X_2)}$ can never be closed for $\gamma' \neq 0$ (we shall come back
to this point in Section IV). As to the difference $X_2 -X_1$,  it becomes large
i.e., $X_2\gg X_1$ due to the presence of the last term in Eq.\,(\ref{eq:58b}) with $D_2>0$, since lowering the
temperature leads also to a decrease in $X_2$.

\subsection{Upper boundary for strength of DM interaction}
In our previous work\ci{ourpart1}, requiring positiveness of self-energies, $X_1$ and $X_2$, we have found a
boundary condition for the strength of EA interaction as $\gamma \leq U|\sigma|$. Now we address the question whether a similar condition be found for the strength of DM interaction $\gamma'$.

First, we note that Eq.\,\re{eq:r0} for $r_0$ has a positive solution regardless the sign (and value) of the parameter $P$. In fact, since $\gamma'>0$, the number of sign changes in this equation is equal to unity, it has exactly one positive solution due to Descartes' Rule of Signs. Hence, in the approximation suggested by Ref.\, \cite{Sirker1},
the right hand side of
Eqs.\,(\ref{eq:Z1Z2}) are positive for any $\gamma'>0$. Thus, when we neglect $\sigma$ and use the
approximation linear in $\gamma'$, there is no upper bound for the strength of DM
interaction, $\gamma'$. However, when we go beyond such an approximation, we have to deal with Eqs.\,\re{eq:X10}
and \re{eq:X20}, where the last terms with $\gamma'^2$ play an important role for large $\gamma'$.
By examining the coefficients of $\gamma'^2$, namely $D_1$ and $D_2$ given in Eqs.\,(2.4) one may find that
$D_2>0$ and $D_1<0$ at any temperature. Now, it can be understood that at large values of $\gamma'$,
the last term in Eq.\,\re{eq:GX2} will dominate over the first and second terms, making the right hand side
of this equation negative. Actual numerical analysis for TlCuCl$_3$ show that this happens at $\gamma'>0.7$\.K
for $H\leq 20$\,T. In reality, $\gamma'$ is rather small: $\gamma'_{optimum}\approx 0.02$\,K (see Section IV).
Anyway, in contrast to  approximation used in Ref.\, \cite{Sirker1}, the present approach, taking into account $\gamma'$ up to the second order, is able to predict an upper bound for the strength of DM interaction
$\gamma'_{max}\approx 0.7$K, beyond which this interaction destroys the condensate of triplons.

\section{Sensitivity of thermodynamic characteristics to anisotropies}
In Ref.\, \cite{ourpart1}  we have shown that the presence of $H_{EA}$ and $D_{DM}$ terms in the bosonic
Hamiltonian with contact interaction may significantly modify the phase and the condensate fraction of BEC. Now we discuss their influence on some physical quantities .

\subsection{ Anomalous density and self-energy}

\begin{figure}[bt]
\begin{minipage}[bt]{0.49\linewidth}
\center{\includegraphics[width=1.1\linewidth]{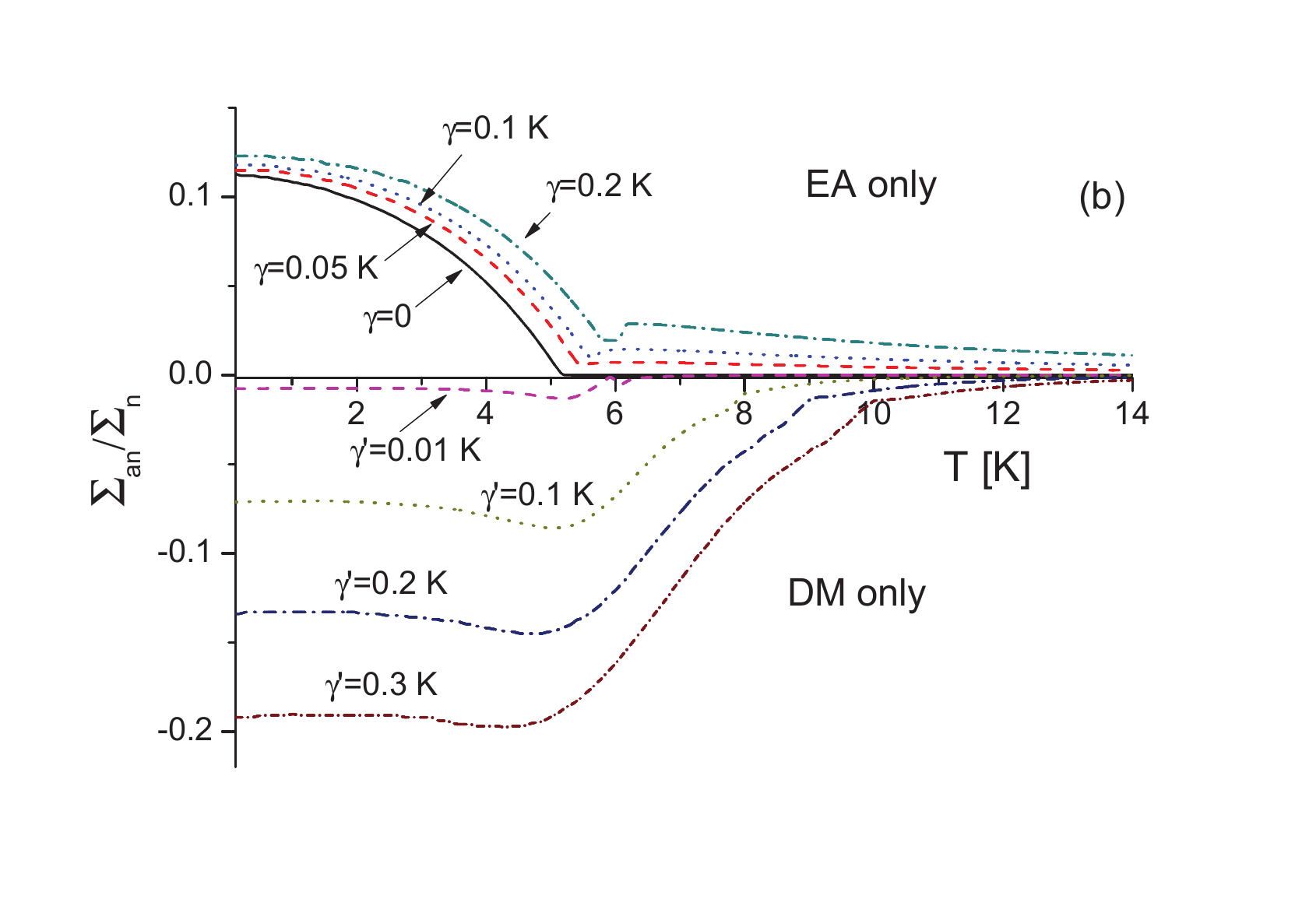} }
\end{minipage}
\hfill
\begin{minipage}[bt]{0.49\linewidth}
\center{\includegraphics[width=1.1\linewidth]{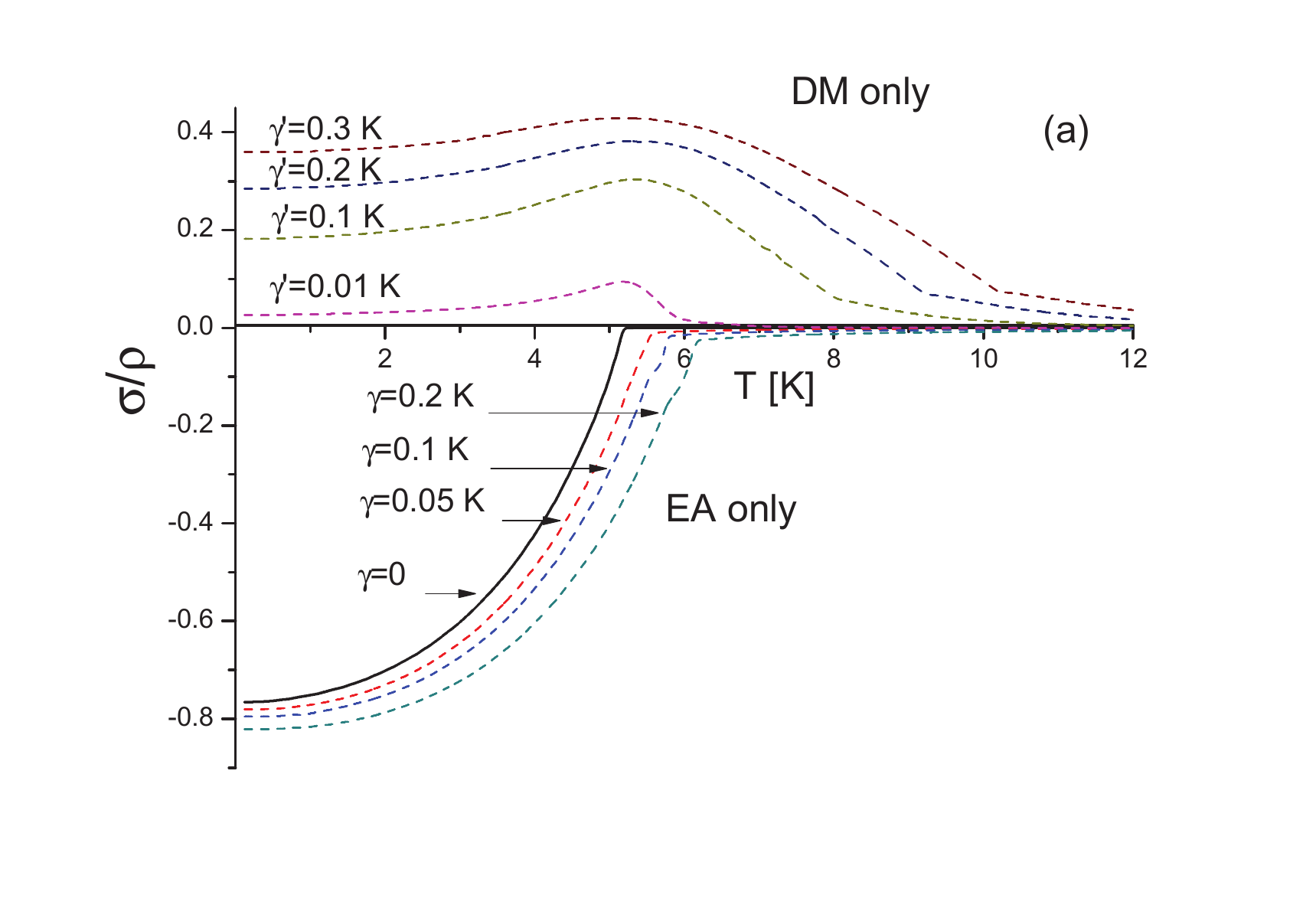} }
\end{minipage}

\caption
{
(a) The ratio of anomalous density $\sigma$ to the total density $\rho$ of triplons
as a function of  temperature for various values of DM
 and EA  interactions. (b) The same as in (a) but for the ratio of
anomalous and normal self-energies. It is seen that the presence of DM interaction
reverses the sign of both anomalous density and self-energy. The solid curves in both figures
correspond to isotropic case with $\gamma=\gamma'=0$. The input parameters are the same
as in Fig.\,\ref{FIGX12RHO0X}.
}
  \label{FIGSIG}
\end{figure}

Firstly we show that even a tiny DM interaction changes the sign of anomalous density $\sigma$, which is
negative for pure BEC in finite systems. In fact, subtracting Eq.\,(\ref{eq:GX2}) from Eq.\,(\ref{eq:GX1}) and using
Eq.\,(\ref{eq:rho1sig}) with $\gamma=0$, one obtains
 \bea
 \sigma=\tilde{S}(\rho_0-\sigma) - \frac{\tilde{S} \gamma'^2 (D_1 + D_2)}{U X_2^2}
 \label{eq:56}
 \eea
where $\bar{S}= U\sum_{k} W_k/E_k $. The formal solution of Eq.\,(\ref{eq:56}) is
\bea
\sigma=\frac{\tilde{S}[\rho_0 -\gamma'^2(D_1 + D_2)/U X_2^2]}{1+\tilde{S}}\, .
 \label{eq:57}
\eea
Now, from the explicit expressions for $D_1$, $D_2$ defined in Eqs.\,(2.4) it can be shown that
$(D_1 + D_2)\leq 0 $. Thus, from Eq.\,(\ref{eq:57}) it is understood that $\sigma(\gamma' \neq 0) \geq 0$ at any temperature for $U>0$, $\gamma'>0$. Numerical results presented in Fig.\,\ref{FIGSIG}(a) confirm
this conclusion. As to the magnitude of anomalous density, it is seen that
both kind of anisotropies lead to increasing of $|\sigma |$, which may reach even
$20 \% $ of the total density of triplons for the moderate values of $\gamma'$.

In Fig.\,\ref{FIGSIG}(b), a similar quantity, namely, the ratio of anomalous self-energy to the normal self-energy,
$\Sigma_{an}/\Sigma_{n}$ is presented.  It is seen that $\Sigma_{an}$ does not vanish even in the normal phase, where it is equal to $\Sigma_{an} {(T>T_c)}=\gamma$. Moreover, the presence of DM interaction changes the sign of $\Sigma_{an}$.

\subsection{Shift in the critical temperature}

The critical temperature $T_c$ is one of the main characteristics of systems undergoing
BEC transition.
It is understood that the presence of any kind of interaction (or  geometry of a trap)
modifies the critical temperature of BEC. Quantitatively this is characterized in
literature by the relative shift of critical temperature $\Delta T_c/T_{c}^{0}$
defined as
\be
\dsfrac{\Delta T_c}{T_{c}^{0}}\equiv \frac{T_c-T_{c}^{0}}{T_{c}^{0}}
\lab{Deltatc}
\ee
where $T_{c}^{0}$ is the critical temperature of BEC transition without the interaction under consideration.

 \begin{figure}[bt]
\begin{minipage}[bt]{0.49\linewidth}
\center{\includegraphics[width=1.1\linewidth]{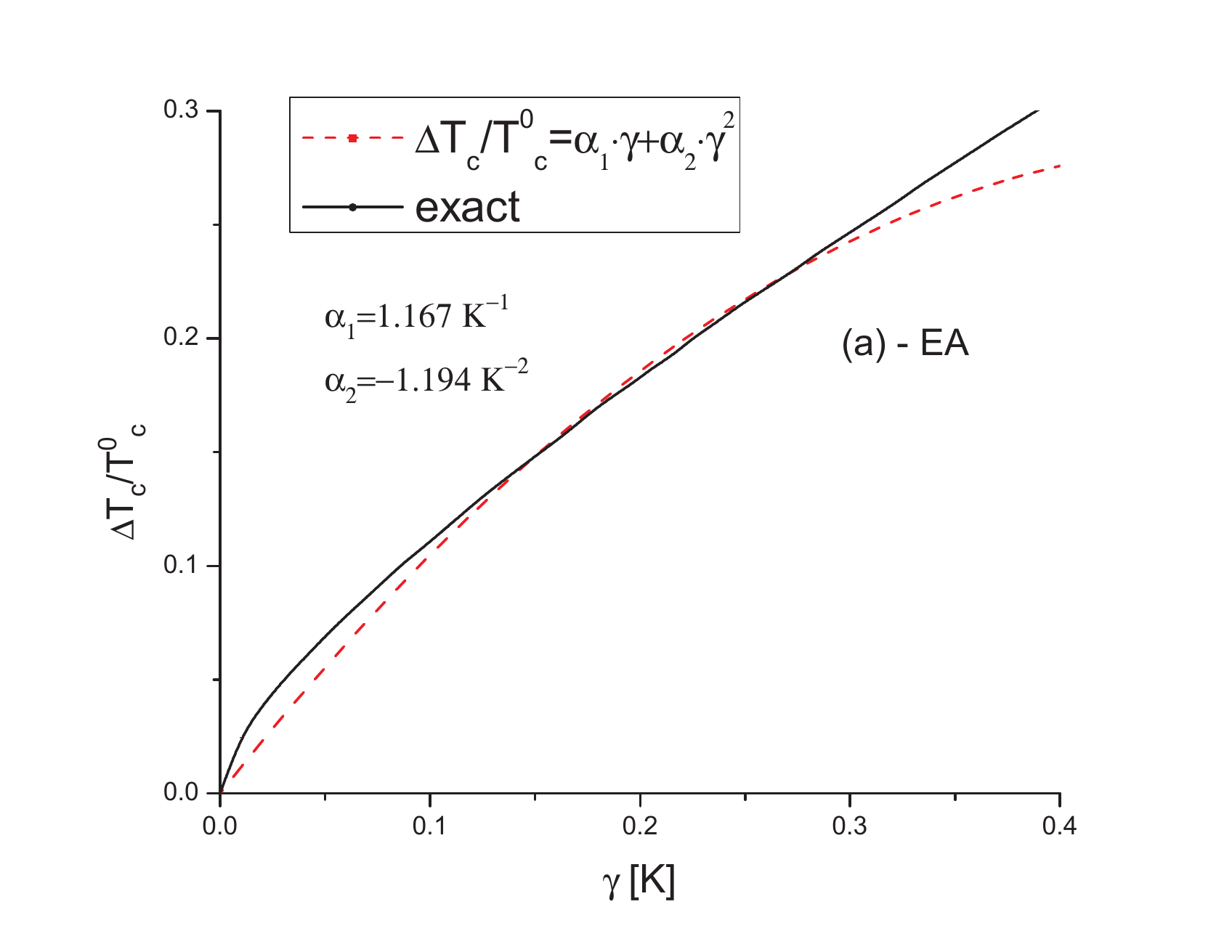} }\\
\end{minipage}
\hfill
\begin{minipage}[bt]{0.49\linewidth}
\center{\includegraphics[width=1.1\linewidth]{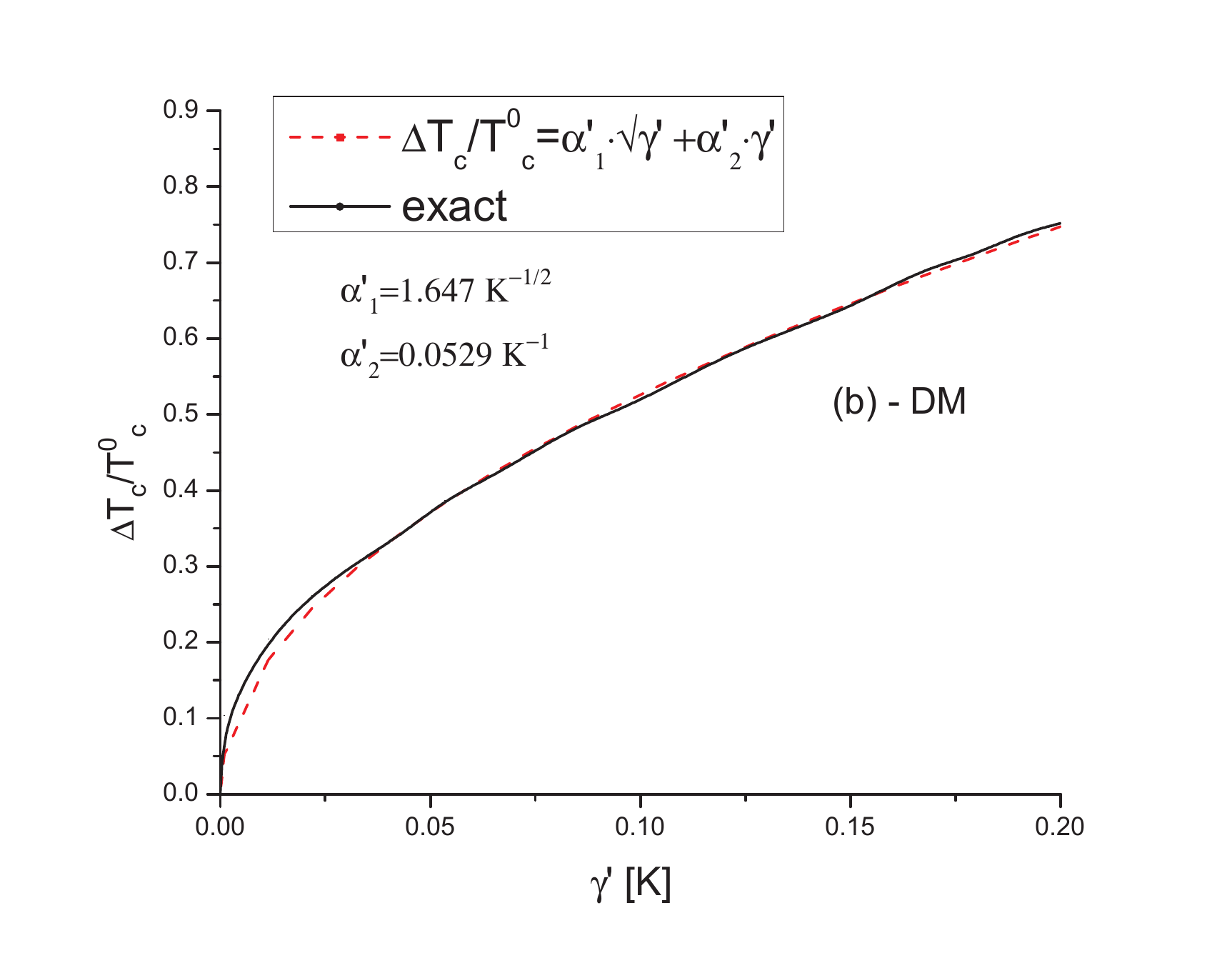} }\\
\end{minipage}

\caption
{
The shift in critical temperature due to EA  (a) and DM (b)
interactions (solid curves). Dashed curves are phenomenological fits.
The input parameters are the same as in Fig.\,\ref{FIGX12RHO0X}.
}
  \label{FIGTC}
\end{figure}

In general, the problem of accurate estimation of the shift turns out to be highly nontrivial,
since close to the phase transition, the physics in the interacting gas is governed by strong fluctuations, which make
perturbation theory inapplicable\ci{yuktc}. Nevertheless, one can find in the literature
some analytical formulas for $\Delta T_c/T_{c}^{0}$ due to interparticle contact interaction\ci{ramos},
due to the trap geometry\ci{arnold}, or due to disorder\ci{ourdisorder,vinokur}.
We now consider how the critical temperature $T_c$ of triplon BEC may be affected by anisotropies.
To find an answer to this question we have to make numerical analysis, since obtaining analytical
estimations turns out to be rather complicated.

In Figs.\,\ref{FIGTC}(a) and (b) we present
the dependence of the shift due to EA and DM interactions, respectively. For weak anisotropies
these can be approximated in powers of $\gamma/U$ and $\sqrt{\gamma'/U}$ as
$\Delta T_c/T_{c}^{0} (\gamma)\approx a_1 (\gamma/U)+a_{2} (\gamma/U)^2$ and
$\Delta T_c/T_{c}^{0} (\gamma')\approx a'_1 \sqrt{(\gamma'/U)}+a'_{2} (\gamma'/U)$
for the cases of EA and DM interactions, respectively. Clearly the optimized parameters
$a_i$ and $a'_i$ depend also on the external magnetic field $H$. Particularly, for TlCuCl$_3$
with $U=315$\,K at $H=8.5$\,T we obtained $a_1/U=1.167$\,K$^{-1}$, $a_2/U^2=-1.194$\,K$^{-2}$,
$a'_1/\sqrt{U}=1.647$\,K$^{-1/2}$ and $a'_2/U=0.053$\,K$^{-1}$, as illustrated in Fig.\,\ref{FIGTC}.

Firstly, one may note that in both cases $\Delta T_c \geq 0$, which means that presence
of the anisotropies shift the critical temperature of BEC transition, (or a crossover in the case of DM anisotropy)
toward higher values. Secondly, it is seen that the influence of anisotropy
is not negligibly small at moderate values of the intensities. For instance, DM interaction
with $\gamma '\approx 0.1$\,K modifies $T_c$ with $\Delta T_c/T_{c}^{0} (\gamma '=0.1$\,K$)\sim$ 50\%.
Thirdly, DM anisotropy modifies the critical temperature more strongly than EA anisotropy. For example, for the
equal values of intensities, say, $\gamma=\gamma'\approx 0.1$\,K, the shift due to DM interaction is nearly five times
larger than due to EA interaction. Thus, the critical temperature is more sensitive to DM interaction
than to EA.

\subsection{Magnetization}

\begin{figure}[bt]
\begin{minipage}[bt]{0.49\linewidth}
\center{\includegraphics[width=1.1\linewidth]{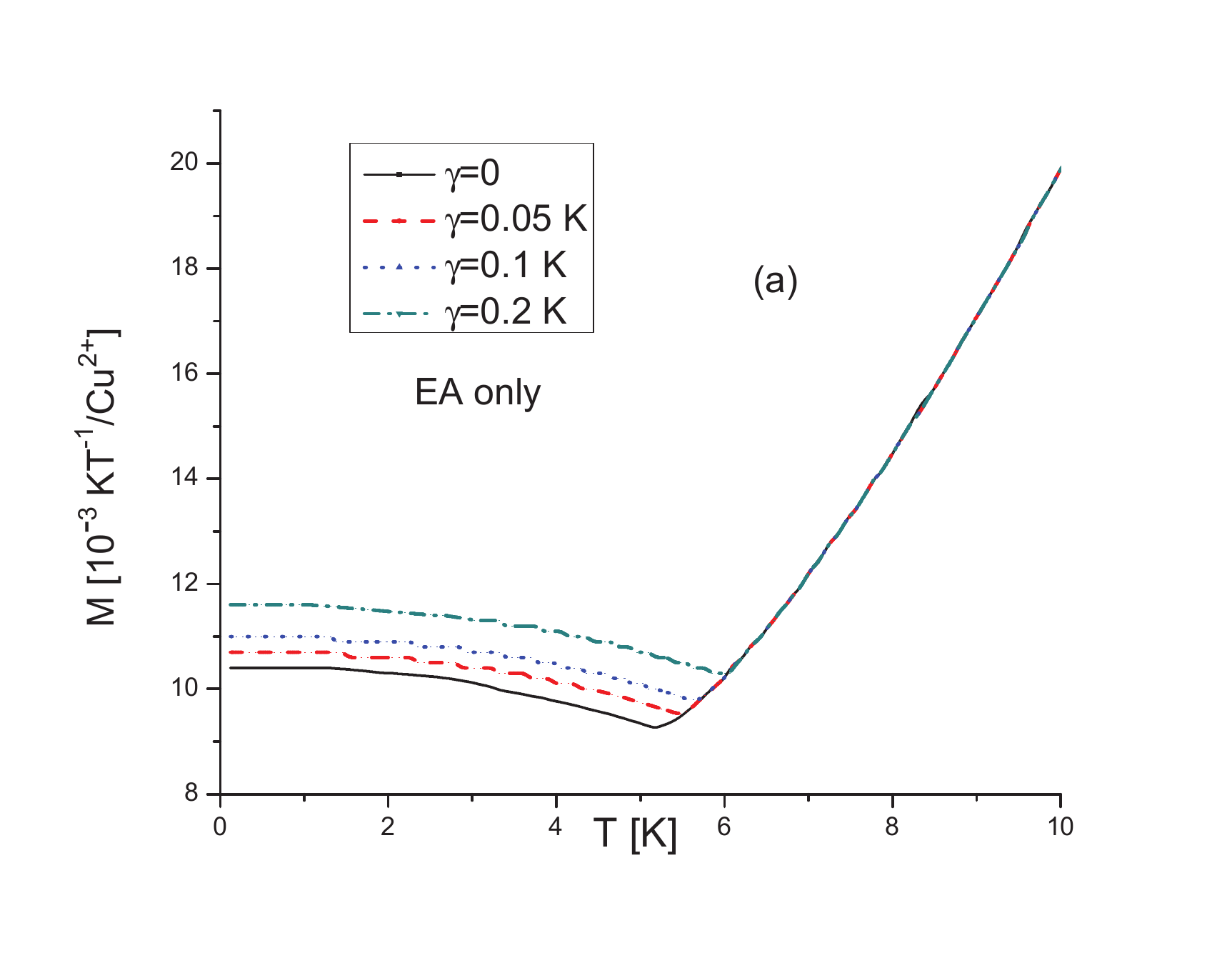} }
\\
\end{minipage}
\hfill
\begin{minipage}[bt]{0.49\linewidth}
\center{\includegraphics[width=1.1\linewidth]{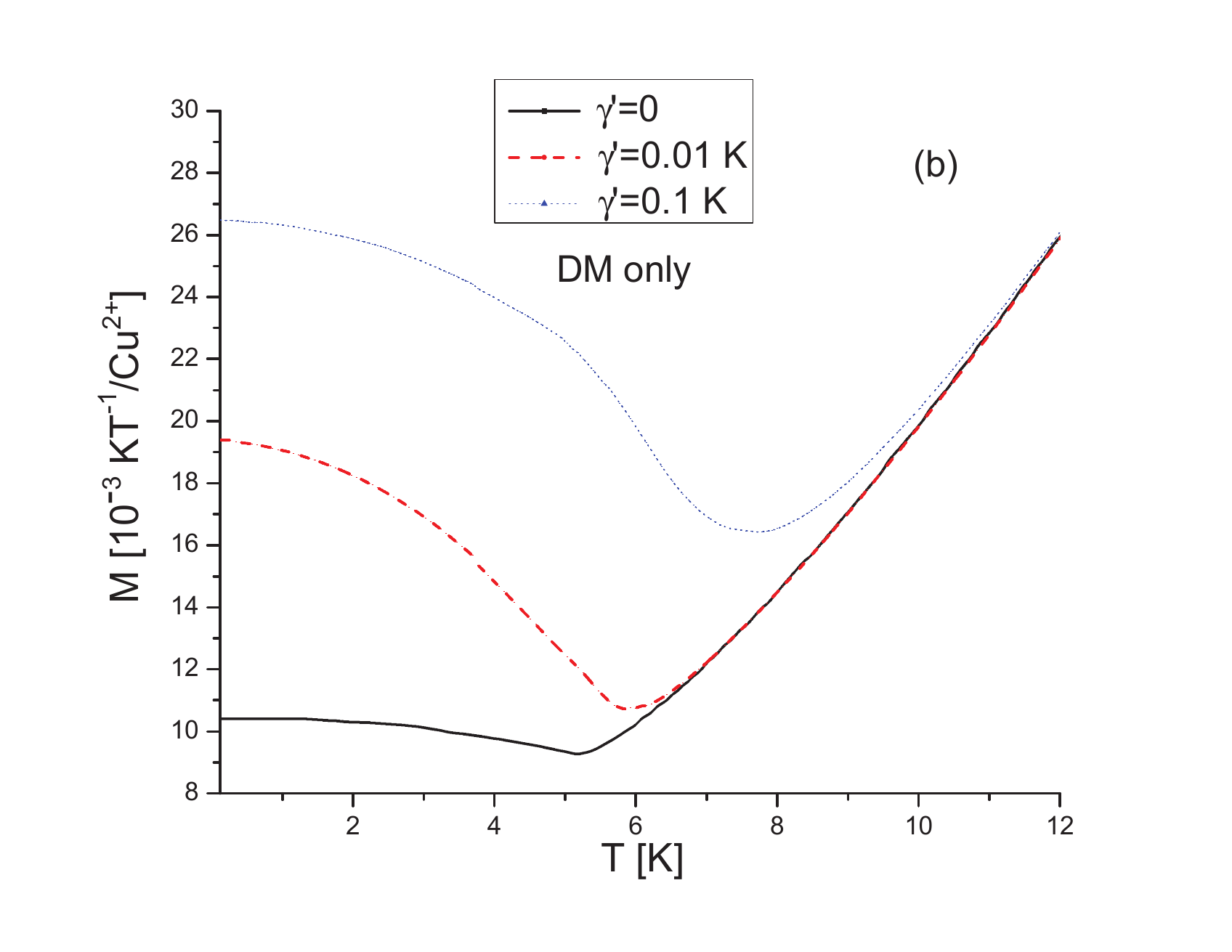} }
\\
\end{minipage}
\medskip
\medskip
\medskip
\begin{minipage}[bt]{0.49\linewidth}
\center{\includegraphics[width=1.1\linewidth]{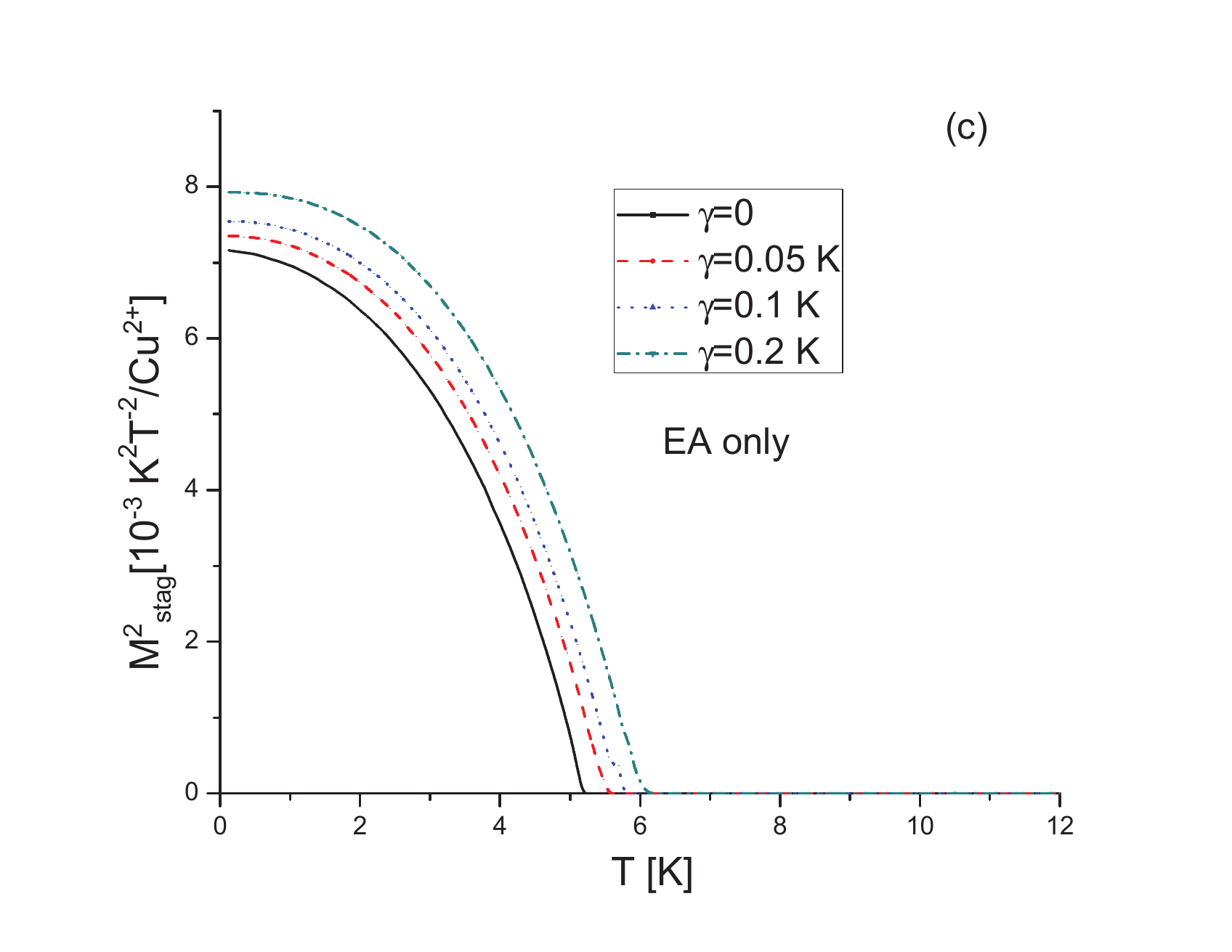}
\\
}
\end{minipage}
\hfill
\begin{minipage}[bt]{0.49\linewidth}
\center{\includegraphics[width=1.1\linewidth]{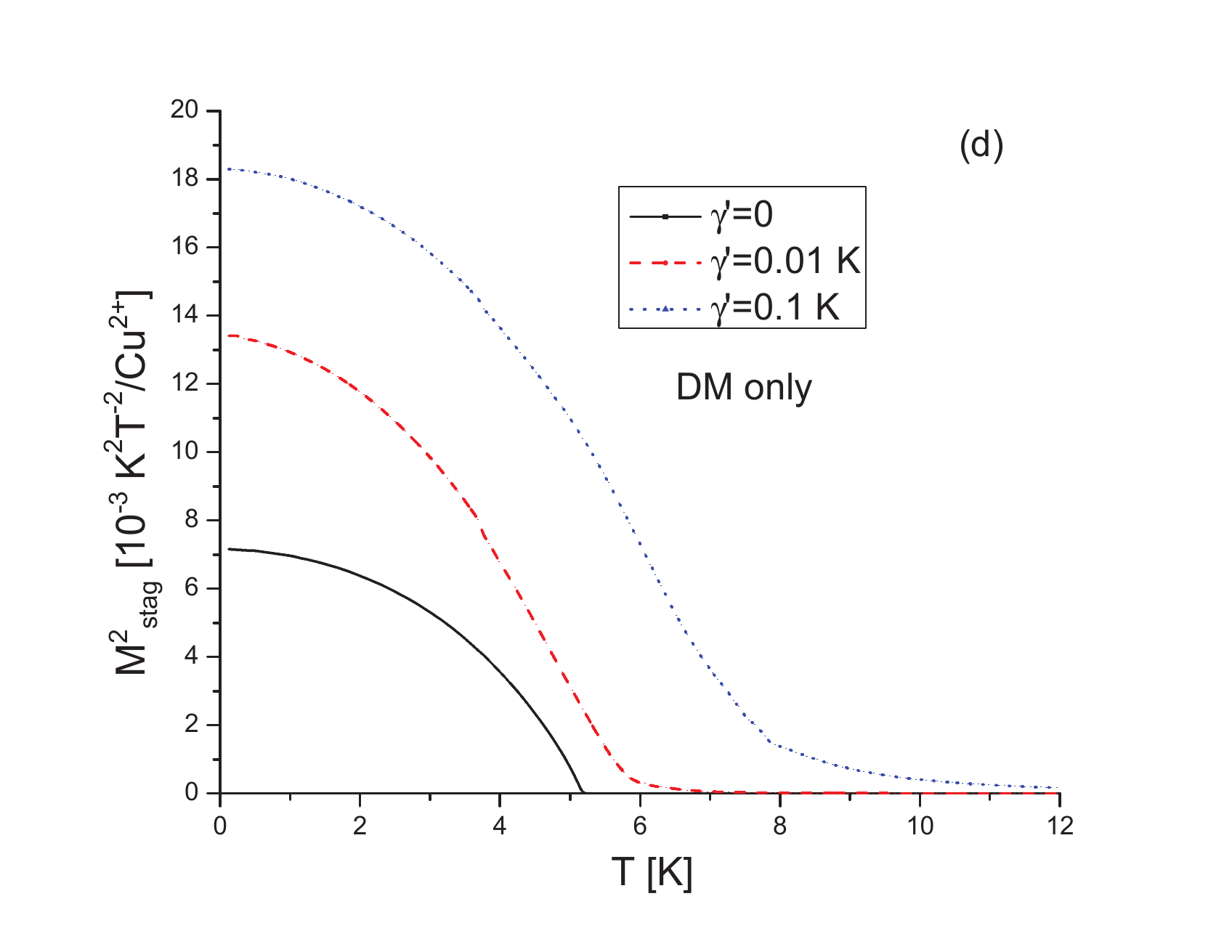}
 \\
}
\end{minipage}
\caption
{
Uniform magnetization as a function of temperature with only EA (a) and DM (b) anisotropies. Solid lines correspond to the isotropic case with $\gamma=\gamma'=0$. (c) and (d) display the square of staggered magnetization
$(M_\bot)^2$. The input parameters are the same as in Figs.\,\ref{FIGX12RHO0X}.
}
  \label{FIGMAG}
\end{figure}

In Figs.\,\ref{FIGMAG} the uniform magnetization $M(T)$ and $M_{\bot}^{2}(T)$ are presented for various values of
$\gamma$ and $\gamma'$ as a function of temperature temperature. It is seen that the EA interaction modifies
both of these quantities mainly at low temperatures ($T\leq T_c$) (Figs.\,\ref{FIGMAG}(a), (c)). As to the DM
interaction its effect is twofold. At low temperatures it enhances $M$ as well as $M_{\bot}$
and in contrast to EA interaction, it prevents the staggered magnetization from vanishing at $T\geq T_c$.
Thus, taking into account of DM anisotropy, at least in the linear form as in Eq.\,\re{Hbde} within MFA, is
inevitable in the accurate description of experimental data on $M_{\bot}$, reported by
Tanaka {\it et al.}\ci{Tanasca2001}.

\subsection{Heat capacity at constant field $C_H$}
In the presence of BEC the heat capacity exhibits the following specific features.
\begin{itemize}
\item
Its dependence on temperature has a well known $\lambda$-shape\ci{huangbook}
which was first observed in superfluid helium\ci{hill}.
\item
Near absolute zero, $C_V(T)$ behaves like $C_V(T)\propto T^3$, due to
a linear energy dispersion, responsible for the superfluidity.
\item
Near the critical temperature $C_V$ has a discontinuity, i.e.,
$\Delta C_V\equiv\ds\lim_{\epsilon\rightarrow 0}
[C_V (T_c-\epsilon )-C_V (T_c+\epsilon )]\neq 0$  which is expected for a second order phase
transition\ci{Landau}.
\end{itemize}
In the present work to study these features of the heat capacity of triplons at constant magnetic field
and in the presence of anisotropies, we evaluate $C_H (T)$ for the case of only EA anisotropy,
(see Fig.5(a))\footnote{$C_H $ in the presence of DM anisotropy will be discussed
in a separate paper.  }.
Firstly, it is seen that in both cases of low and high temperatures, behavior of $C_H(T)$
is not modified significantly, almost coinciding with
the case without anisotropy (solid lines in Fig.\,5(a)). That is the anisotropies are prominent mainly
in the critical region. Further,  EA interaction leaves the famous $\lambda$-shape almost
unchanged (Fig.\,5(a)). Actually, in the presence of EA anisotropy, there is a definite point $T_c$
where $\rho_0(T=T_c)=0$, which separates BEC and normal phases. This leads to a sharp
maximum in the specific heat (see Fig.\,5(a)), as in the case of a pure BEC
without any anisotropy (solid line in Fig.\,5(a)).
 \begin{figure}[bt]
\begin{minipage}[bt]{0.49\linewidth}
\center{\includegraphics[width=1.1\linewidth]{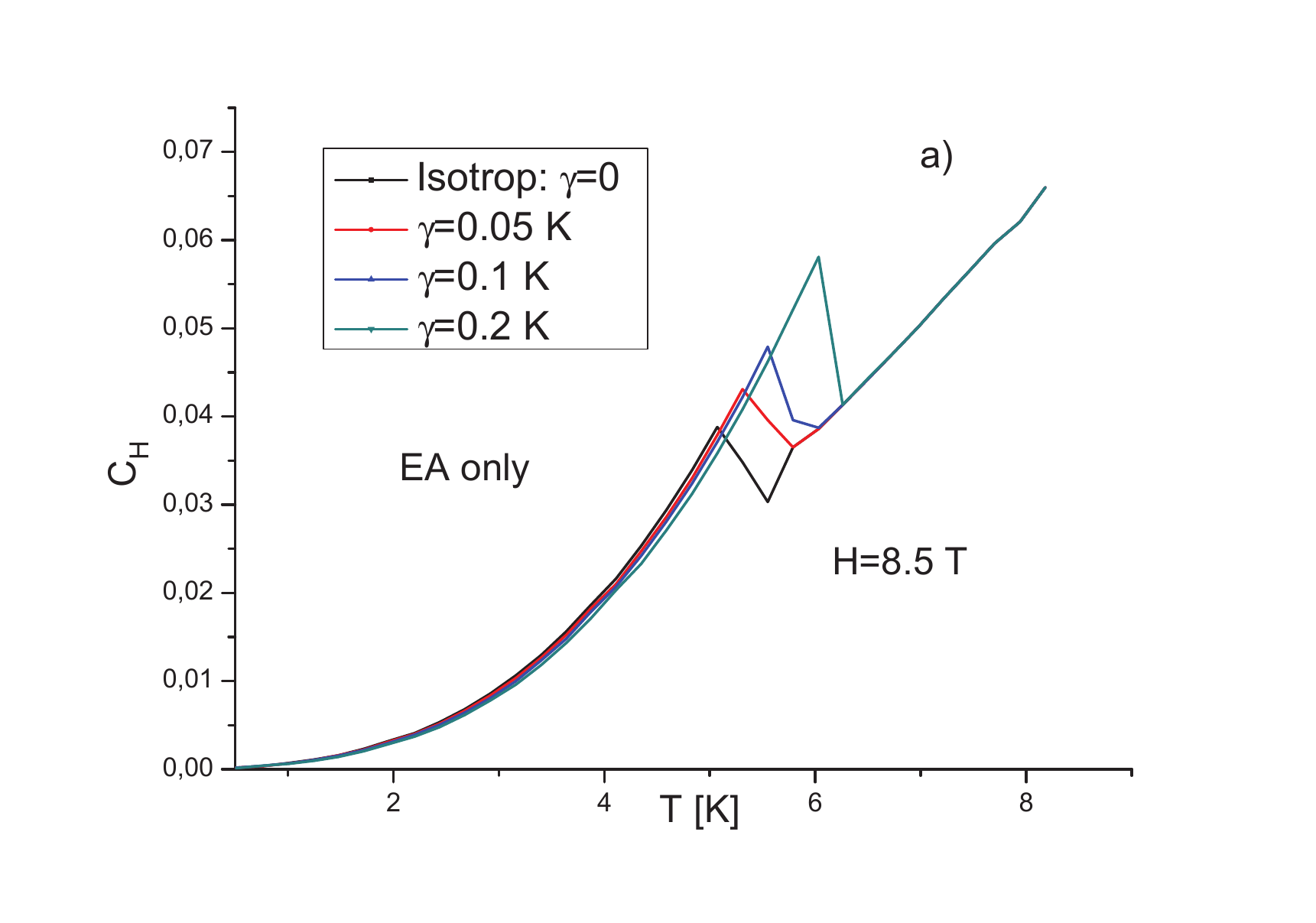} }
\end{minipage}
\hfill
\begin{minipage}[bt]{0.49\linewidth}
\center{\includegraphics[width=1.2\linewidth]{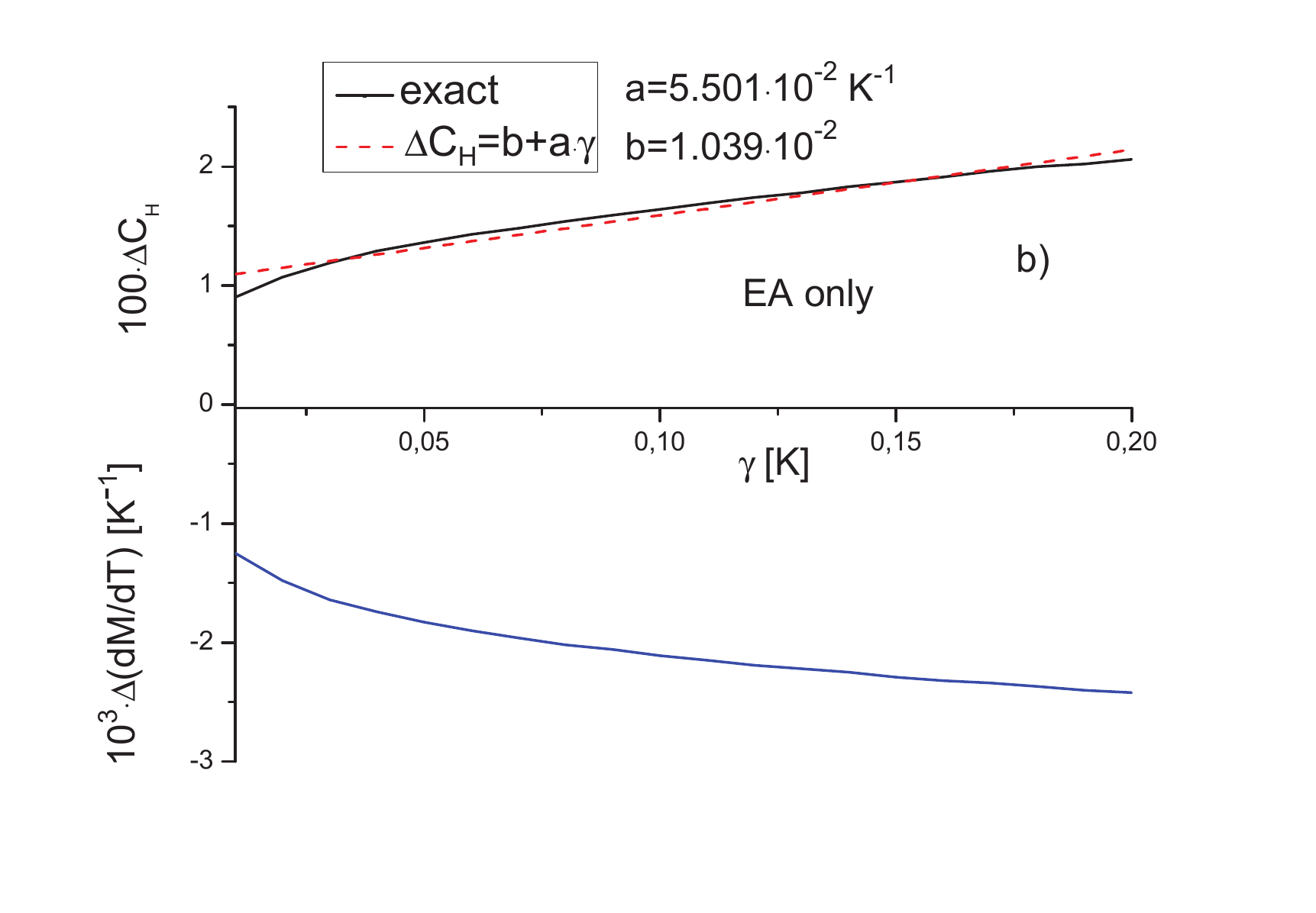} }
\end{minipage}
\hfill
\caption
{The heat capacity $C_H$ as a function of temperature with only  EA
anisotropy  (a). Solid lines correspond to the isotropic case
with $\gamma=\gamma'=0$. (b): The discontinuity in $C_H$ (upper panel) and $d\rho/dT$ (lower panel)
near $T=T_c$.
 The dashed curve is a phenomenological fit.
 The input parameters are the same as in Figs.\,\ref{FIGX12RHO0X}. Here one should note that
 the presence of anisotropies modifies not only $C_H$ but also $T_c$.}
  \label{CH}
\end{figure}
In order to find
the shift in $\Delta C_H$ due to the EA interaction we evaluated $\Delta C_H$
as a function of the strength of EA interaction using Eqs.\,\re{eq:CH}, \re{setdrho} and \re{drhotbec}.
The results are presented in Fig.\,5(b). It is seen that $\Delta C_H(\gamma=0, \gamma'=0)\approx 0.01$
i.e., the discontinuity is positive for a pure BEC, as it is expected\ci{huangbook,ouriman}.
For small values of the EA strength, $0<\gamma\leq 0.1$\,K,
the function $\Delta C_H(\gamma)$ can be approximated (dashed line in Fig.\,5\,(b)) by
$\Delta C_H(\gamma)\sim b+a{\gamma}$. For example, at $H=8.5$\,T, the optimal values are:
$b\approx 0.01$ and $a=0.055$\,K$^{-1}$.
In spite of the presence of EA anisotropy, $\Delta C_H$ remains finite which proves that
the corresponding BEC-like transition may be classified as a second order phase transition.
From Fig. 5(b) it is seen that, $\Delta C_H(\gamma)$ is always positive (upper panel),
while $\Delta (dM/dT)$ remains negative for any $\gamma$. This is in a good agreement with
Ehrenfest relation\ci{ourmce}
\be
\Delta C_H=-\left\{T\left(\dsfrac{\partial H}{\partial T}\right)[\Delta \left(\dsfrac{\partial M}{\partial T}\right)]\right\}|_{T=T_c}.
\lab{erenfest}
\ee

\section{Results for realistic parameters for $\hbox{TlCuCl}_3$ and discussions}

In the previous section we studied the effect of anisotropies on thermodynamic quantities. Particularly,
we have shown that in contrast to EA interaction, DM interaction modifies their behavior dramatically.
It smears BEC transition to  a crossover and changes the sign of the anomaleous density.
 Clearly, the significance, or measurability of such effects depend on their  interaction strengths $\gamma$ and $\gamma'$. Evidently, unless we have realistic values for these parameters
for a real material, our studies will remain purely academic.

Among the 3D quantum dimerized magnets with a spin gap TlCuCl$_3$ seems to be the most
experimentally studied
compound\cite{shermansound,matsumotoprl,matsumotoPRB,misguich,glazkov,delamoreexper,Ruegg,oosawacv,oosawa99,rueggnature,oosawaprb65,rueggprl,cavadiniprb63,cavadiniprb65,rueggappphys2002,kimura2007}.
The observation of a finite $M_{\bot}$ at $T\geq T_c$\ci{Tanasca2001}
uniquely indicates the presence of DM interaction with a finite $\gamma'$. Thus, using existing experimental
data on the magnetization and the heat capacity of TlCuCl$_3$, we have made an attempt
to obtain optimal values of  input parameters of the present approach. The result for $H // b$ is as follows.
$g=2.06$, $U=367$\,K, $\gamma=0.05$\,K and $\gamma'=0.0201$\,K. The magnetizations $M$
and $M_{\bot}$ corresponding to this set of parameters are depicted in Figs.\,6(a) and (b), respectively.
It is seen that the inclusion of DM anisotropy gives a good description
of the staggered magnetization especially at higher temperatures (see, inset of Fig.\,6(b)). Moreover,
taking into account the anomalous density $\sigma$ leads to a better description of
$M$ e.g., at low temperatures, compared with approximation suggested in Ref.\, \cite{Sirker2},
where $\sigma$ has been neglected.

\begin{figure}[bt]
\begin{minipage}[bt]{0.49\linewidth}
\center{\includegraphics[width=1.1\linewidth]{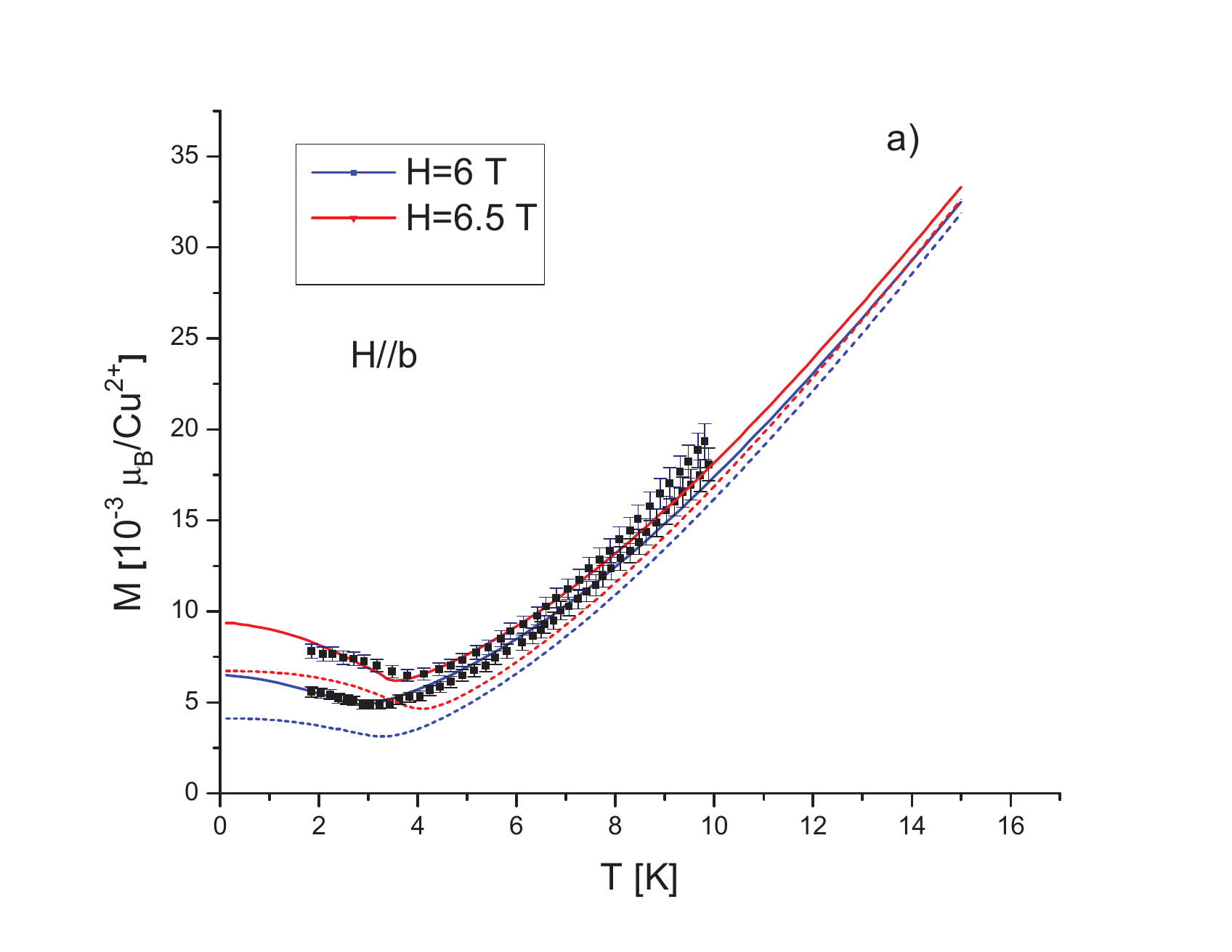} }
\\
\end{minipage}
\hfill
\begin{minipage}[bt]{0.49\linewidth}
\center{\includegraphics[width=1.1\linewidth]{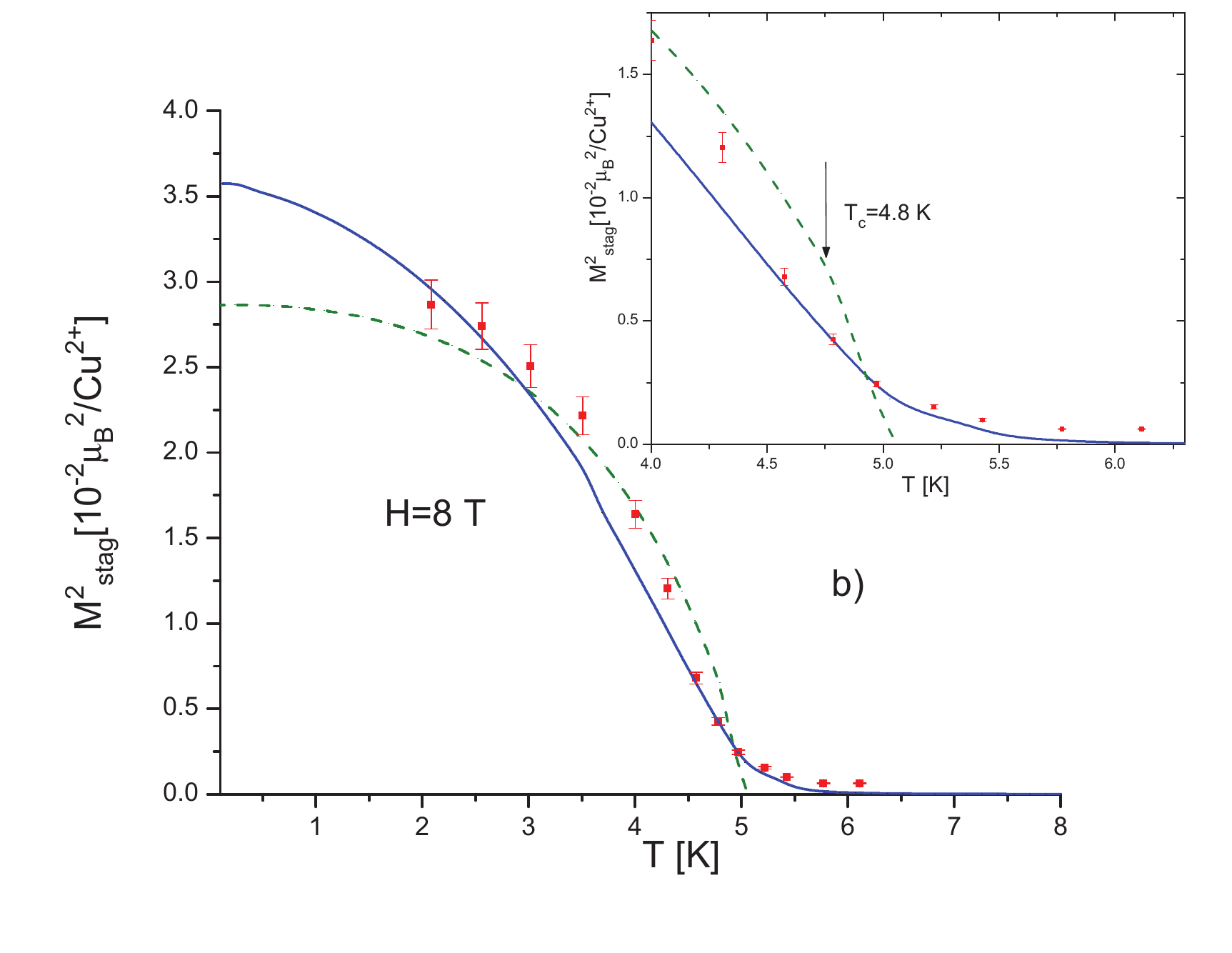} }
\\
\end{minipage}
\hfill
\caption{
Uniform (a) and staggered (b) magnetizations for TlCuCl$_3$,  $H // b$. Solid and dashed lines correspond to the present approximation and approximation in Ref. [28], respectively. Experimental data are taken from
Ref. [34]. The optimized parameters are $\gamma=0.05$\,K, $\gamma'=0.0201$\,K and
$U=367$\,K. Diamagnetic and other contributions to total experimental magnetizations are taken into account following the ansatz by Dell'amore {\it et al.} Ref. [40].
}
  \label{MAG}
\end{figure}
Remarkably, the experimental fit of parameters can be reached with rather small values
of anisotropies, namely $\gamma/U= 1.36\times10^{-4}$
and $\gamma'/U= 5.47\times 10^{-5}$. In order to compare $C_H$ with existing experimental data,
one needs to perform calculations
in the presence of both kinds of anisotropies and solve the problem concerning the  extraction
of a phonon contribution from experimental curves. This rather complicated task will be the subject of
our separated paper.

 Thus we have found that, the experimental data on magnetization
 of TlCuCl$_3$ can be well described by the present approach. On the other
hand there exist experimental measurements on the energy of magnetic excitations.
In the following subsection we shall compare our results with these experiments.

 \subsection{Energy dispersion}

As it has been outlined in the Introduction, a spin gapped quantum magnet e.g., TlCuCl$_3$
has a dimer structure and a finite energy gap at zero field $\Delta_{ST}$ between the singlet
$S=0$ ground state and the first excited states $S=1$. When an external field is applied and reaches
a critical value $H_c=\Delta_{ST}/g\mu_B$ the gap is closed due to the Zeeman effect, as it is illustrated in Fig.\,7(a).
The excitation spectrum of this compound so far was studied in detail by inelastic
neutron scattering (INS)\ci{rueggnature,oosawaprb65,rueggprl,cavadiniprb63,cavadiniprb65,rueggappphys2002}
as well as ESR measurements\ci{glazkov,kimura2007}.

The INS studies confirmed that the system becomes quantum critical at $H_c\approx 5.7$\,T
where the energy of the lowest Zeeman-split excitation $| 1,-1\rangle$ crosses the nonmagnetic ground state
$| 0,+0 \rangle$. Above this lowest mode the system remains in a gapless Goldstone mode and develops a linear
dependence on the momentum, which is a good signal of occurrence of BEC.
On the other hand, ESR study on this compound gave evidence for a tiny spin gap
with minimal value $\Delta_{an}\sim 0.2$\,meV, which was not observed in INS experiments (see Fig.\,7(a)).
Therefore, the experimental situation on the energy spectrum of TlCuCl$_3$ has not been totally
clear. In fact, on the one hand, the lowest excitation spectrum for $H_c\leq H \leq H_{saturation}$
at $T\leq T_c$ is gapless, $\Delta_{an}(INS)=0$, on the other hand, it has a finite gap
$\Delta_{an}(ESR) \neq 0$ and hence can not be linear.  Theoretically, it is clear that if the gap remains
finite it may be caused by a lattice anisotropy.
\begin{figure}[bt]
\begin{minipage}[bt]{0.49\linewidth}
\center{\includegraphics[width=1.1\linewidth]{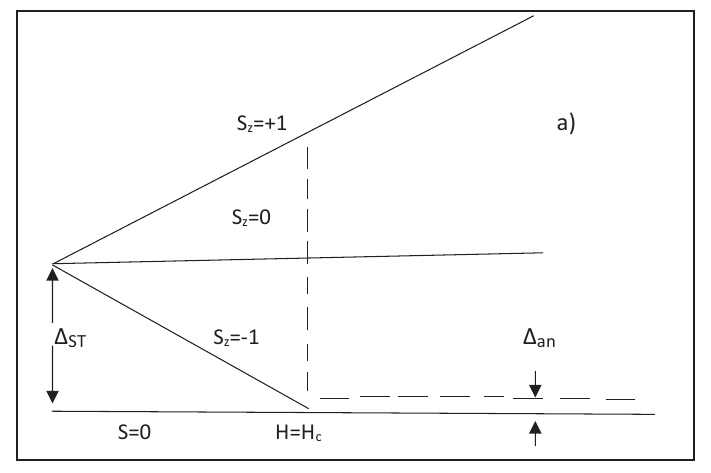} }
\\
\end{minipage}
\hfill
\begin{minipage}[bt]{0.49\linewidth}
\center{\includegraphics[width=1.1\linewidth]{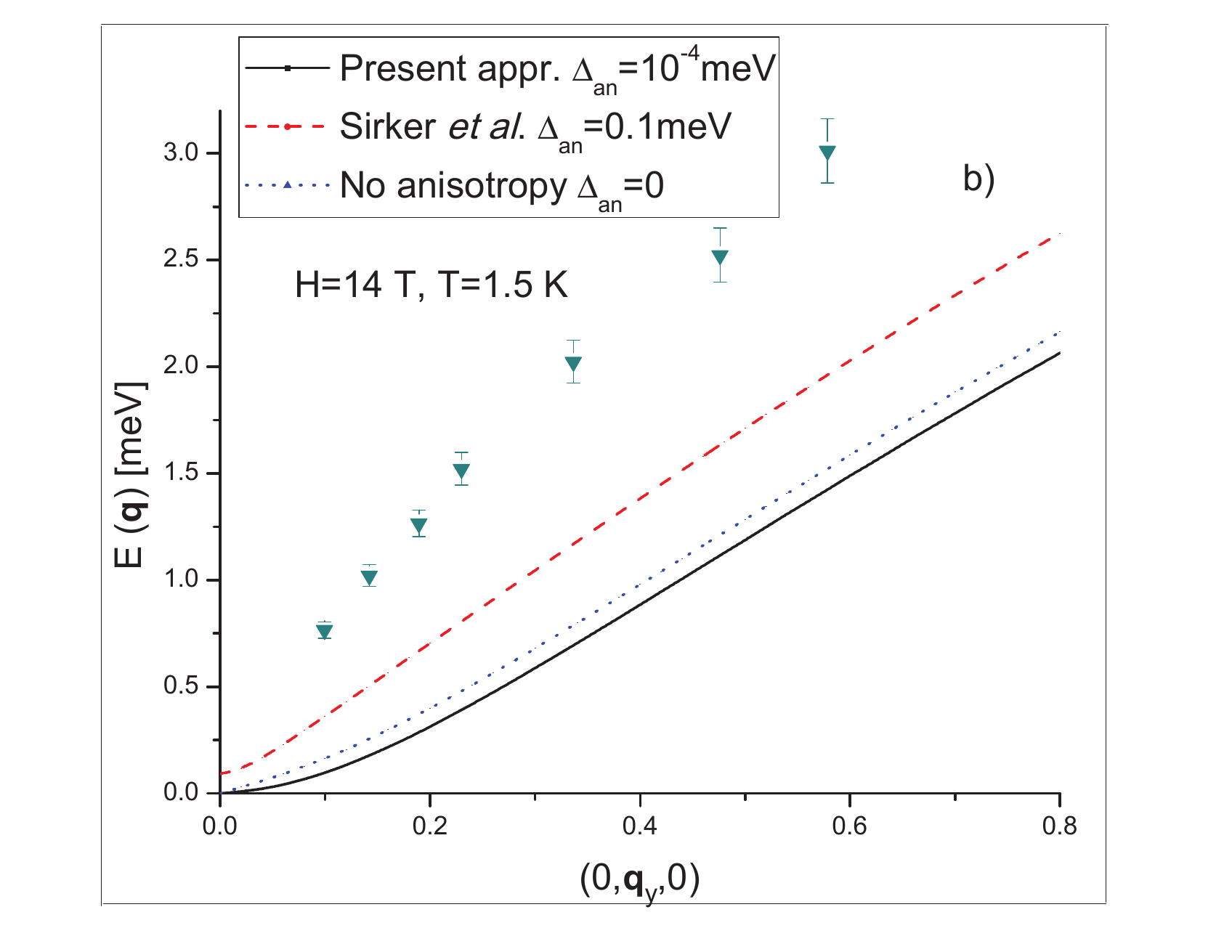} }
\\
\end{minipage}
\hfill
\caption{
(a) The schematic illustration of energy levels of a spin-gapped system. At $H=H_c$ the gap $\Delta_{ST}$ closes and may reopen due to anisotropies with a tiny gap $\Delta_{an}$.
(b) Energy dispersion of the low-lying magnetic excitations in TlCuCl$_3$. The solid,  dashed, and dotted lines correspond to the present approximation including anisotropies; approximation by Ref. [28]
and without anisotropy, respectively. The experimental data are taken from Ref. [44].
}
  \label{DIS}
\end{figure}
Here for clarity, it should be noted that in the present version of mean-field theory one should distinguish two
types of energy dispersions. A bare dispersion $\veps_k\sim k^2/2m$
and the dispersion of collective excitations, given as  $E_k=\sqrt{\veps_k+X_1}\sqrt{\veps_k+X_2}$, where the
self-energies $X_1$ and $X_2$ are discussed in Section II. The dispersion of elementary excitations at zero
field $\veps_k$ is well studied experimentally\ci{oosawaprb65,cavadiniprb63} and
presented as a function of momentum and intra (inter)-dimer interactions $J_i$ as $\veps_k(J_i)$. One can find
in the literature an explicit expression for $\veps_k(J_i)$ with its optimized parameters\ci{matsumotoprl,misguich,cavadiniprb63}, which has also been used in the present work with the normalization
$\veps_k\mid _{k\rightarrow 0}=\vec{k}^2/2m $\ci{ourmce}.

As to the energy spectrum at $H\geq H_c$, it is clearly model dependent. For example, in the isotropic case
for $T\leq T_c$ it is gapless, given by $E_k=\sqrt{\veps_k+X_1}\sqrt{\veps_k}\sim ck +\mathcal{O}(k^3)$, thus, $\Delta_{an}=E_k\mid _{k\rightarrow 0}=0$ in agreement
with experimental results by R\"{u}egg {\it et al.}\ci{rueggnature}. In the presence of anisotropies it has
a finite gap $\Delta_{an}=\sqrt{X_1 X_2}$, where $X_1$ and $X_2$  are defined by Eqs.\,(2.12) and (2.13).
Using our optimal input parameters we obtained a finite but rather small
value  $\Delta_{an}(H=14$\,T, $T$=1.5\,K)=10$^{-4}$\,meV, which is consistent with INS measurements,
but not with ESR: $\Delta_{an}(H=14$\,T, $T$=1.5\,K)=0.2\,meV\ci{glazkov,kimura2007}.
In Fig.\,7(b) we present quasiparticle spectrum $E_k=\sqrt{\veps_k+X_1}\sqrt{\veps_k+X_2}$, $(k_x=k_z=0, k_y=\pi q_y)$ for $H=14$\,T at $T=1.5$\,K. It is seen that, the excitation energy in the present approximation
is almost linear, in accordance with experimental results.  However, the experimental values of $E_{k}^{exp}$
are rather underestimated. This can be understood as follows. As it has been shown  in
Section II at low temperatures the self-energies, especially $X_1$ is rather small (Fig.\,1). Our input parameters optimized by experimental magnetizations
lead to much smaller values: $X_1(H=14$\,T, $T$=1.5\,K)=$0.67\times10^{-5}$\,K,
$X_2(H=14$\,T, $T$=1.5\,K)=0.19\,K, thus
$X_1 \ll X_2$. As a result, the momentum dependence of the dispersion is similar to that of isotropic one, $E_k=\sqrt{\veps_k+X_1}\sqrt{\veps_k+X_2}\sim \sqrt{\veps_k}\sqrt{\veps_k+X_2}$
which is practically nothing but the Goldstone mode. Thus, we may come to the conclusion that
in accordance with present approximation
the lowest excitation energy of TlCuCl$_3$ at very low temperatures has a rather small, but finite gap and
exhibits, practically,  a linear dispersion at small momentum, in spite of the presence of EA and DM interactions.
Note that, a similar  situation has been observed for compounds Sr$_3$Cr$_2$O$_8$ and Ba$_3$Cr$_2$O$_8$
which have DM interaction, but no anisotropy gap, i.e., $\Delta_{an}($Sr$_3$Cr$_2$O$_8$)=0,
$\Delta_{an}($Ba$_3$Cr$_2$O$_8$)=0.\ci{wangprb2014}

  \subsection{ Discussions}

   In the present section, having fixed the parameters of the theory by magnetization data on
   TlCuCl$_3$, we have studied its energy spectrum above the critical field at $T\leq 1.5$\,K. We have found that
   the description of magnetizations for $H//b$ is quite good, while that of the energy dispersion of the low-lying magnetic excitations needs to be improved. In some sense, this brings to mind the situation in nuclear
   physics: one can choose optimal parameters for the nucleon-nucleon potential by experimental data on cross sections, but fails to accurately describe the binding energies of light nuclei. Anyway,
   the main reason of our failure seems to be the simplicity of the Hamiltonian $H_{DM}$ used here (the last term
   in Eq.\,(1.2)). In fact, in deriving this linear Hamiltonian it has been assumed that, the DM vector
   is parallel to $x$, i.e., $\vec{D}=[D_x,0,0]$\ci{Sirker1}. Therefore, it is naturally expected that
   by using a more general form for $H_{DM}$, where other components of $\vec{D}$
   are also included\ci{Miyahara,matsumoto2008} one will be able to describe
   not only magnetizations, but also excitation energies in the extended version of the present
   mean-field approach. Note that, by neglecting the other components of the DM vector, one cannot
   describe magnetizations for $H\bot (1,0,\bar{2})$ either.

\section{Summary and Conclusions}

We have studied effects of lattice anisotropies on thermodynamic characteristics of spin-gapped quantum
magnets for $H_c\leq H < H_{Saturation}$ by applying our extended mean-field based approach, proposed
in our previous work\ci{ourpart1}. This nonperturbative approach takes into account the anomalous density and
both EA and DM interactions more accurately than it is done e.g., in the HFP approximation.
We derived explicit expressions for some thermodynamic quantities which include the self-energies $X_1$
and $X_2$, and the condensate fraction $\rho_0$. Analysis of the coupled equations with respect to
these three quantities show that at high temperatures $T\gg T_c$, the self-energies $X_{1,2}$
are not significantly affected by EA and DM interactions. Meanwhile, the latter strongly modifies
the condensate fraction converting BEC transition into a crossover.

At low temperatures the DM interaction increases $\rho_0$, but leads to rather small values of $X_{1}$,
compared with the isotropic
case. As a result, the energy dispersion  $E_k=\sqrt{(\veps_k+X_1)(\veps_k+X_2)}$, develops
a linear dependence at small momentum, in accordance with experimental measurements.

In contrast to EA interaction, the presence of DM interaction, even in the simple linear form
in the Hamiltonian, modifies the anomalous density, changing its sign.
 Particularly, it is expected that,  the  usual ``$\lambda $-shape" of the heat capacity
 disappears
due to strong DM interactions.
On contrary, the presence of only EA anisotropy
leaves the ``$\lambda $-shape" of the heat capacity  unchanged.
The discontinuity in $C_H$  close to the critical temperature
is shifted significantly for
moderate values of the intensity of the exchange anisotropy.

We have found optimal input parameters of the Hamiltonian  for the compound
TlCuCl$_3$ which describes experimental data on magnetizations, at least for $H// b$, quite well.
This set of parameters lead to a linear dispersion of energy of quasiparticles, but predicts
a fairly small value of an anisotropy gap, estimated by ESR measurements.

In future work we plan to extend our Hamiltonian by taking into account a more realistic DM interaction
to obtain better description of experimental data on
the spectrum of low lying excitations, as well as the heat capacity.

\section*{Acknowledgements}
We are indebted to Andreas Schilling for useful discussions and comments. AR acknowledges support
by TUBITAK-BIDEB (2221), AK is supported by the Ministry of Innovative Development of the Republic of Uzbekistan
and thankful to group of J. Osterwalder at Physics Institute of University of Zurich. BT is supported by Science
and Technological Council of Turkey (TUBITAK) under Grant No: 119N689 and Turkish Academy of Sciences
(TUBA) under Grant No. AD21. This work is partly supported by funding from Academy of Sciences of the
Republic of Uzbekistan.

\newpage
\appendix

\section{ Explicit expressions for some thermodynamic parameters}


\numberwithin{equation}{section}

\setcounter{equation}{0}

As shown in Section II the physics of the cases with and without
anisotropies are quite different. In the presence of DM interaction all useful expressions for physical observables
may be found by setting $\xi=i$ in Eqs.\,(\ref{eq:X10}) and (\ref{eq:X20}) which are to be solved with the restrictions $X_1\geq0$, $X_2 \geq 0$. However, when DM is absent ($\gamma'=0, \gamma \neq0$), one must be aware of the Hugenholtz-Pines (HP) theorem\cite{HP} which holds in the limit $\gamma \rightarrow 0$. Below, we discuss  these two cases separately.

\subsection{Mode 1: $\gamma'=0, \gamma\neq 0$ }

We start from the explicit expression for $\Omega$,
\bea
\Omega(\gamma'=0, \gamma \neq 0, \xi=1)&= U\rho_1^2 +\frac{U(\sigma^2+\rho_0^2)}{2} + \rho_1 \left(-\frac{X_1}{2}-\frac{X_2}{2}-\mu +2U\rho_0\right) +\nn
& \sigma \left(\frac{X_2}{2}-\frac{X_1}{2}+\gamma +U\rho_0\right) + \gamma \rho_0   -\mu_0 \rho_0 + \Omega_T
\label{eq:B1}
\eea
where
\begin{subequations}
	\begin{align}
\Omega_T&= \frac{1}{2}\sum_k (E_k - \veps_k) + T \sum_k \ln (1-e^{-\beta E_k})\label{eq:Bomega}\\
X_1&=U(3\rho_0 + 2\rho_1 +\sigma) -\mu +\gamma   \label{eq:BX1}\\
X_2&=U(\rho_0 + 2\rho_1 -\sigma) -\mu -\gamma \label{eq:BX2}\\
E_k&=\sqrt{{(\veps_k +X_1)(\veps_k +X_2)}}
\end{align}
\end{subequations}
and $\mu_0=2U\rho_1 +U\sigma +\gamma +U\rho_0$ is introduced  to avoid the Hohenberg-Martin dilemma\ci{hohmartin} in the condensate phase. In this phase, $\rho_0 (T\leq T_c)=0$ and
HP relation may be written in a slightly ``broken" form\cite{23 our aniz}:
\bea
\Sigma_{n} -\Sigma_{an} -\mu=X_2=2\gamma
\label{eq:sigan}
\eea
which gives a gapless energy dispersion in the $\gamma\rightarrow0$ limit: $E_k(T<T_c)=\sqrt{(\veps_k +X_1)(\veps_k +2\gamma)}$. Using (\ref{eq:BX1}), (\ref{eq:BX2}) and (\ref{eq:sigan}) yields
\be
\ba
X_1(T\le T_c)=2U\sigma + 2U\rho_0 +4\gamma , \quad X_2(T\le T_c)=2\gamma
\lab{x64}
\ea
\ee
whose solution is positive definite due to $|\rho_0|\geq |\sigma|$. This equation may be rewritten in a more convenient form as
\bea
\Delta=\frac{X_1}{2}=\mu+2U(\sigma -\rho_1) +5\gamma
\label{eq:B5}
\eea
where $\sigma$ and $\rho_1$ are given by Eqs.\,(\ref{eq:17a}) and (\ref{eq:rho1sig}) with $X_2=2\gamma$, $X_1=2\Delta$ and $\mu=g\mu_B(H-H_c)$. Having solved Eq.\,(\ref{eq:B5}) with respect to $\Delta$ one may evaluate the densities as
\begin{subequations}
	\begin{align}
\rho_0 = \frac{\Delta-2\gamma-U\sigma}{U}\, ,\\
\rho=\frac{\Delta+\mu+\gamma}{2U}\, .
\lab{rhototbec}
\end{align}
\end{subequations}
In the normal phase ($T>T_c$), one may neglect $\rho_0$ in Eqs.\,(\ref{eq:BX1}) and (\ref{eq:BX2}) to obtain
\begin{subequations}
	\begin{align}
X_{1,2}(T>T_c)&=2U\rho-\mu \pm \gamma \pm\sigma \label{eq:B7a}\\
\rho(T>T_c)&=\sum_{k}\frac{1}{e^{\beta \omega_k} -1}   \label{eq:B7b} \\
\omega_k&=\sqrt{(\veps_k +X_1)(\veps_k +X_2)}\, .
\label{eq:B7c}
\end{align}
\end{subequations}

The entropy $S$, heat capacity $C_H$, and Gr{\"u}neisen parameter may be found as\ci{ourjt,ourcharak}
\begin{subequations}
	\begin{align}
S&=-\left(\frac{\partial \Omega}{\partial T}\right)_H = - \sum_{k} \ln\left[1-exp{(-\beta \mathcal{E}_{k})}\right]+\beta \sum_{k}\frac{\mathcal{E}_{k}}{e^{\beta \mathcal{E}_{k}}-1} \label{eq:S} \\
C_H&= T\left(\frac{\partial S}{\partial T}\right) =\frac{1}{4} \sum_{k} W_k' \mathcal{E}_{k}(\mathcal{E}_{k,T}'-\beta \mathcal{E}_{k}) \label{eq:CH}\\
\Gamma_H&= -\frac{1}{C_H}\left(\frac{\partial \mu}{\partial T}\right)_H=\frac{g\mu_B}{C_H}\left(\frac{\partial \rho}{\partial T}\right)
\label{eq:GH}
\end{align}
\end{subequations}
where $\mathcal{E}_k=\sqrt{(\veps_k +X_1)(\veps_k +X_2)}$, $\mathcal{E}_{k,T}'=({\partial \mathcal{E}_{k}}/{\partial T})_H$
and $X_{1,2}$ are given by Eqs.\,\re{x64} and \re{eq:B7a}.
Below we give explicit expressions for $\mathcal{E}_{k,T}'$ and  $\rho_T'$ for normal ($T>T_c$) and BEC  ($T\leq T_c$) phases where the critical temperature is defined at the point
$\rho_0(T=T_c)=0$.

\subsubsection{Critical temperature and density }
The condition $\rho_0(T=T_c)=0$ leads the following coupled equations with respect to  $T_c$ and
$\sigma_c$\ci{23 our aniz}:
\be
\ba
\dsfrac{\mu}{2U}+\dsfrac{\sigma_c+3\gamtil}{2}-\sum_k\dsfrac{f_b(E_{k}^{c})}
{E_{k}^{c}}[\veps_k+U(\sigma_c+3\gamtil)]=0\\
\\
\sigma_c+U(\sigma_c+\gamtil)\sum_k\dsfrac{f_b(E_{k}^{c})}{E_{k}^{c}}=0
\ea
\lab{tcsigc}
\ee
where ${E_{k}^{c}}=E_k(T\rightarrow T_c)=\sqrt{\veps_k+X_{1}^{c}}\sqrt{\veps_k+2\gamma}$,
$X_{1}^{c}=2U(\sigma_c+2\gamtil)$, $f_b(x)=1/(\exp(x/T_c)-1)$   and $\gamtil=\gamma/U$.
The critical density is given by
\be
\rho(T=T_c)=\dsfrac{\mu}{2U}+\dsfrac{\sigma_c+3\gamtil}{2}\equiv\rho_c\, .
\lab{rhocgam}
\ee
\subsubsection{{Normal phase}}
For $T>T_c$, differentiating Eq.\,(\ref{eq:B7b}) and using Eq.\,(\ref{eq:B7c}) we obtain
following set of equations:
\be
\ba
\rho_T'(T>T_c)=\dsfrac{b_1 a_{22}-b_2 a_{12} }{a_{11}a_{22}-a_{12}a_{21}     },
\quad \sigma_T'(T>T_c)=\dsfrac{b_2 a_{11}-b_1 a_{21} }{a_{11}a_{22}-a_{12}a_{21}     },
\\
\\
\dsfrac{d\omega_k}{dT}=\dsfrac{U}{\omega_k}
[2(\veps_k-\mu+2U\rho)\rho_T'-U(\gamtil+\sigma)\sigma_T'    ],
\\
\\
a_{11}=1-2A+\dsfrac{U}{2}\sum_k\dsfrac{(\veps_k-\mu+2U\rho)^2(-2+4W_k-\omega_k W_k')     }
{\omega_{k}^{3}    },\\
\\
a_{12}=\dsfrac{U^2}{4}\sum_k\dsfrac{(\veps_k-\mu+2U\rho)(\sigma+\gamtil)(2-4W_k+\omega_k W_k')     }
{\omega_{k}^{3}    },\\
\\
a_{22}=1-\dsfrac{U^3}{4(1+A)^2}\sum_k\dsfrac{\gamtil(\sigma+\gamtil)(2-4W_k+\omega_k W_k')     }
{\omega_{k}^{3}    },\\
\\
a_{21}=\dsfrac{2\gamtil a_{12}}{(1+A)^2(\sigma+\gamtil)    },\quad b_1=-\dsfrac{1}{4T} \sum_k (\veps_k-\mu+2U\rho)W_k' ,\\
\\
 b_2=\dsfrac{U\gamtil}{4T(1+A)^2} \sum_k W_k' , \quad A=U\sum_k\dsfrac{1}{\omega_k (\exp(\beta\omega_k)  -1 )},
\lab{setdrho}
\ea
\ee
where $W_k'=-\beta/ \sinh^2(\beta \omega_k/2)$ and $W_k=1/2 \coth (\beta \omega_k/2)$.

\subsubsection{BEC phase}
In this case $\mathcal{E}_k(T)=E_k(T)=\sqrt{(\veps_k +2\Delta(T))(\veps_k +2\gamma)}$  differentiation of which gives
\bea
\mathcal{E}_{k,T}'=E_{k,T}'=\frac{\veps_k+2\gamma}{E_k} \Delta_T'\, .
\eea
To find $\Delta_T'=\left({\partial \Delta}/{\partial T}\right)$, we can differentiate both sides of Eq.\,(\ref{eq:B5}) with respect to $T$ and solve it for $\Delta_T'$. The result is
\bea
\Delta_T'=\left(\frac{\partial \Delta}{\partial T}\right)_H =\frac{US_3}{2T(2S_4 +1)}
\eea
with $S_3=\sum_k W_k'(\veps_k +2\Delta)$ and $S_4=U\sum_k {(4W_k + E_k W_k')}/{4E_k}$. As to $\rho_T'$ it can be found directly from \re{rhototbec} as
\be
\rho_T'(T\leq T_c)=\dsfrac{S_3}{4T(2S_4+1)}.
\lab{drhotbec}
\ee
At last, setting in Eqs.\,\re{setdrho} and \re{drhotbec} $\rho_0=0$, $T=T_c$, $E_k=\omega_k=
\sqrt{\veps_k+X_{1}^{c}}\sqrt{\veps_k+2\gamma}$ one may define the cusp in $\rho_T'$ as
$\Delta \rho_T'=\rho_T' (T_c^-)-\rho_T' (T_c^+)   $ presented in Fig.5b. As to the  cusp in $C_H$, presented
also in Fig.5b may be found in a similar way from Eq.s \re{eq:CH}, \re{setdrho} and \re{drhotbec}.

\subsection{Mode 2: $\gamma' \neq 0, \gamma\neq0$ case}
The expressions for $S$, $C_H$, and $\Gamma_H$ remain formally unchanged. However, explicit expressions for
$\mathcal{E}_{k,T}'$ and $\partial\rho/ \partial T$ in Eq.\,(\ref{eq:CH}) and (\ref{eq:GH}) are quiter complicated.
Implicit differentiation of $E_k=\sqrt{(\veps_k +X_1(T))(\veps_k +X_2(T))}$ gives
 \bea
 \mathcal{E}_{k,T}'=\frac{\partial \mathcal{E}_{k}}{\partial T}=\frac{(\veps_k +X_2)X_1'+(\veps_k +X_1)X_2'}{2E_k}
 \eea
where $X_1'=\partial X_1/ \partial T$ and $X_2'=\partial X_2/ \partial T$ whose explicit expressions
will be given below.
Now, differentiating both sides of  equations $\rho_1=(A+B)/2$ and $\sigma=(B-A)/2$
with respect to temperature one obtains
\begin{subequations}
\begin{align}
\frac{d\rho_0}{dT}=C_\rho[X_1'(B_1'+3A_1')+X_2'(A_1'+3A_2')+A_t'+B_t']
\\
\frac{d\rho_1}{dT}=\frac{X_1'}{2}(A_1'+B_1')+\frac{X_2'}{2}(A_1'+A_2')+\frac{1}{2}(A_t'+B_t')
\\
\frac{d\sigma}{dT}=\frac{X_1'}{2}(B_1'-A_1')+\frac{X_2'}{2}(A_1'-A_2')+\frac{1}{2}(B_t'-A_t')
\\
\frac{d\rho}{dT}=\frac{d\rho_0}{dT}+\frac{d\rho_1}{dT}
\label{eq:18}
\end{align}
\end{subequations}
where $C_{\rho}=-U\rho_0^{3/2}/(\gamma'+2U\rho_0^{3/2})$, $A_t'=-({\beta}/{4})\sum_{k} W_k'(\veps_k+X_1)$,
$B_t'=-({\beta}/{4})\sum_{k} W_k'(\veps_k+X_2)$ and $A_i'=\partial A/\partial X_i$, $B_i'=\partial B/\partial X_i$
given in Eq.\,(\ref{a123}).

In the above equations $X_1'=dX_1/dT $ and $X_2'=dX_2/dT$ are still unknown. To find them we rewrite Eq.\,(29a) and (29b) in our previous paper\cite{ourpart1} in the following equivalent form
\bea
M_{11}A_1'+M_{12}B_1'=0 \label{eq:16N}
\\
M_{12}A_1'+M_{11}A_2'+\frac{2\gamma'^2}{X_2^2}=0
\label{eq:16M}
\eea
with $M_{11}=-X_2+2U\rho_0+\gamma'/\sqrt{\rho_0}$, $M_{12}=-X_1+2U\sigma+2\gamma+\gamma'/\sqrt{\rho_0}$ .

Now, differentiating both sides of Eq.\,(\ref{eq:16N}) and(\ref{eq:16M}) and solving the resulting equations for $X_1'$and $X_2'$ one finally gets
\bea
X_1'=\frac{A_{12}b_2-b_1A_{22}}{A_{11}A_{22}-A_{21}A_{12}}, \quad X_2'=\frac{A_{21}b_1-b_2A_{11}}{A_{11}A_{22}-A_{21}A_{12}}
\label{eq:19N}
\eea
where
\begin{subequations}
\begin{align}
A_{11}=A_1'M_{11,1}+M_{11}A_{11}''+B_1'(-1+M_{12,2}')+M_{12}B_{11}''
\\
A_{12}=A_1'(-1+M_{11,1}')+M_{11}A_{12}''+M_{12,2}'B_1'+M_{12}A_{11}''
\\
A_{21}=A_1'(-1+M_{12,2}')+M_{12}A_{11}''+M_{11,1}'A_2'+M_{11}A_{12}''
\\
A_{22}=A_1'M_{12,2}'+M_{12}A_{12}''+(-1+M_{11,1}')A_2'+M_{11}A_{12}''
\\
b_1=M_{11,t}'A_1'+M_{11}A_{1,t}''+M_{12,t}'B_1'+M_{12}B_{1,t}''
\\
b_2=M_{12,t}'A_1'+M_{12}A_{1,t}''+M_{11,t}'A_2'+M_{11}A_{2,t}''
\end{align}
\end{subequations}
and we have introduced the abbreviations $M_{ij,k}'=\partial M_{ij}/\partial X_k$ and $A_{ij}''=\partial^2A/\partial X_i \partial X_j$. $f_t'$ is an explicit derivative with respect to temperature $f_t'(\varphi(X_1(T),X_2(T),T)=df/dT-(\partial f/\partial\varphi)X_1'-(\partial f/\partial\varphi)X_2'$.

\end{document}